\documentstyle[amsmath,amssymb,aps,prb,epsfig]{revtex}
\twocolumn
\renewcommand{\vec}{\bf}
\newcommand{\bj}{{\cal B}_J}
\newcommand{\F}{{\cal F}}
\newcommand{\E}{{\cal E}_J}
\newcommand{\rot}{{\rm \nabla\times}}
\newcommand{\hj}{{{\vec h}_J}}
\newcommand{\hp}{{\vec h}_p}
\newcommand{\sh}{{\rm sinh}}

\newcommand{\ch}{{\rm cosh}}
\newcommand{\avu}{{\overline{U}_{44}}}
\newcommand{\avc}{{\overline{C}_{44}}}
\newcommand{\Hm}{H^{\rm m}_c}
\newcommand{\Ha}{H_{ab}}
\newcommand{\Ll}{\overset{\leftrightarrow}{\Lambda}}
\begin{document}
% for two column  activate the line below...
\twocolumn[\hsize\textwidth\columnwidth\hsize\csname@twocolumnfalse\endcsname
\draft
\title{The London theory of the crossing-vortex lattice in highly anisotropic layered superconductors}
\author{S.E. Savel'ev\cite{byline1}, J. Mirkovi\'c\cite{byline},
and K. Kadowaki}
\address{Institute of Materials Science, The University of Tsukuba, 1-1-1 Tennodai, Tsukuba 305-8573, Japan, and \\ CREST, Japan Science and Technology Corporation (JST), Japan}
\maketitle

\begin{abstract}
A novel description of Josephson vortices (JVs) crossed by the
pancake vortices (PVs) is proposed on the basis of the anisotropic
London equations. The field distribution of a JV and its energy
have been calculated for both dense ($a<\lambda_J$) and dilute
($a>\lambda_J$) PV lattices with distance $a$ between PVs, and the
nonlinear JV core size $\lambda_J$. It is shown that the
``shifted'' PV lattice (PVs displaced mainly along JVs in the
 crossing vortex lattice structure), formed in high
 out-of-plane magnetic fields $B_z>\Phi_0/\gamma^2 s^2$
{[A.E. Koshelev, Phys. Rev. Lett. {\bf 83}, 187 (1999)]}, transforms into
 the PV lattice ``trapped'' by the JV sublattice at a certain field,
 lower than $\Phi_0/\gamma^2s^2$, where $\Phi_0$ is the flux quantum, $\gamma$ is the
 anisotropy parameter and $s$ is
 the distance between CuO$_2$ planes.
 With further decreasing $B_z$, the
free energy of the crossing vortex lattice structure ( PV and JV
sublattices coexist separately) can exceed the free energy of the
tilted lattice (common PV-JV vortex structure) in the case of
$\gamma s<\lambda_{ab}$ with the in-plane penetration depth
$\lambda_{ab}$ if the low ($B_x<\gamma\Phi_0/\lambda_{ab}^2$) or
high ($B_x\gtrsim \Phi_0/\gamma s^2$) in-plane magnetic field is
applied. It means that the crossing vortex structure is realized
in the intermediate field orientations, while the tilted vortex
lattice can exist if the magnetic field is aligned near the
$c$-axis and the $ab$-plane as well. In the intermediate in-plane
fields $\gamma\Phi_0/\lambda_{ab}^2\lesssim B_x \lesssim
\Phi_0/\gamma s^2$, the crossing vortex structure with the
``trapped'' PV sublattice seems to settle in until the lock-in
transition occurs since this structure has the lower energy with
respect to the tilted vortex structure in the magnetic field
${\vec H}$ oriented near the $ab$-plane. The recent experimental
results concerning the vortex lattice melting transition and
transitions in the vortex solid phase in
Bi$_2$Sr$_2$CaCu$_2$O$_{8+\delta}$ single crystals are discussed
in the context of the presented theoretical model.
\end{abstract}
\pacs{PACS numbers: 74.60.Ge, 74.60.Ec, 74.72.Hs}
\tighten

% for two column  activate the line below...
\vskip.2pc]

% body of paper here
\narrowtext
The mixed state of high temperature superconductors is complex and
rich with various vortex phases \cite{blatter,brandt}. Besides the
vortex lattice described by 3D anisotropic Ginzburg-Landau model,
the new types of vortex structures can occur within the large part
of the phase diagram of the mixed state where the coherence length
along the $c$-axis is smaller than the distance between CuO$_2$
planes. In such a case, the magnetic field, aligned with the
$c$-axis, penetrates a superconductor in the form of quasi
two-dimensional pancake vortices (PVs) \cite{clem} while the field
applied parallel to the $ab$-plane generates Josephson vortices
(JVs) in the layers between CuO$_2$ planes \cite{bul,feinberg}. In
magnetic fields tilted with respect to the $c$-axis, PVs and JVs
can form a common tilted lattice \cite{feinberg} or exist
separately as a crossing (combined) lattice
\cite{ledvij,koshelev}. The tilted lattice represents the inclined
PVs stacks in fields applied close to the $c$-axis  while, at
higher angles, the pieces of JVs linking PVs are developed
\cite{feinberg,ledvij}. The crossing lattice is another structure
containing both a PV stack sublattice and a JV sublattice which
coexist separately.

The vortex-solid phase diagram in the tilted magnetic fields was
first proposed by Bulaevskii, Ledvij, and Kogan \cite{ledvij}.
According to their model, which does not take into account the
interaction between PV and JV sublattices in the crossing lattice
structure, the tilted lattice is formed for all orientations of
the magnetic field until the lock-in transition \cite{lock-in}
occurs if the in-plane London penetration depth $\lambda_{ab}$ is
larger than the Josephson vortex core with size $\gamma s$
($\gamma$ is the anisotropy parameter and
 $s$ is the distance between CuO$_2$ planes). In the opposite limit,
$\gamma s>\lambda_{ab}$, the tilted lattice transforms into the
crossing lattice (as the magnetic field is inclined away from the
$c$-axis) at a certain angle before the lock-in transition happens
\cite{ledvij}. Later, the possibility of the coexistence of two
vortex sublattices with different orientations was analyzed
numerically by comparing the free energy of such system with the
free energy of mono-oriented tilted vortex lattice at different
field orientations and different absolute values of the external
magnetic field for the case of 3D anisotropic (London model)
superconductors \cite{brandt-sudbo,crossing-plate,crossing-london}
as well as layered (Lawrence-Doniach model \cite{LD})
superconductors \cite{crossing-LD}. According to
 that analysis \cite{crossing-london,crossing-LD} performed for
 $\gamma = 50 - 160$, the crossing lattice can be energetically preferable
 in the quite low magnetic fields ($B=\sqrt{B_z^2+B_x^2}\lesssim\Phi_0/\lambda_{ab}^2$)
  in the intermediate field
 orientations $0<\theta_1<\theta<\theta_2<\pi/2$ with $\theta=\arctan(B_x/B_z)$
 ($B_z$ and $B_x$ are the field component along the $c$-axis and parallel to the $ab$-plane,
 respectively). However, the interaction of two
 coexisting vortex sublattices  was not considered in those works
 \cite{brandt-sudbo,crossing-plate,crossing-london,crossing-LD}.
  Recently, Koshelev \cite{koshelev} has studied the case of
extremely anisotropic superconductors $\gamma s\gg \lambda_{ab}$
and has shown that the crossing lattice can occupy substantially
larger region of the vortex lattice phase diagram in the oblique
fields due to the renormalization of the JV energy $\E$ through
the interaction of a Josephson vortex and the
 PV sublattice. In addition, such interaction leads to the
 attraction of PVs to JVs \cite{huse,koshelev} at low out-of-plane
 magnetic fields $B_z$ (the some sort of pinning effect).
 This pinning may induce transitions between different
 substructures of the crossing lattice structure. However, there is still no theoretical investigation
 how PV sublattice can influence on the JV lattice in the crossing
 vortex structure in the case of moderate anisotropic
  superconductors with $\gamma s<\lambda_{ab}$.
 In this regard, the phase diagram \cite{ledvij} of the vortex
 lattice for strongly anisotropic layered superconductors
  should be reconsidered (at least for the case of $\gamma s<\lambda_{ab}$)
  by taking into account the renormalization of $\E$ and the
  pinning of PVs by JVs.

The vortex structures in highly anisotropic layered
superconductors are usually studied on the basis of the nonlinear
discrete Lawrence-Doniach model \cite{LD}, but this model is quite
complex and the detailed analysis of the vortex system is
complicated. On the other hand, the layerness of superconductors
can be ignored on scales larger than the size of the nonlinear
Josephson vortex core. Therefore, the linear anisotropic London
model could be applied for a study of the vortex crossing lattice
outside JV cores. In this paper we introduce the extended London
theory which allows to describe the crossing lattice as well as to
calculate the energy $\E$ and the field distribution of JV in the
presence of the crossed PV sublattice. It is shown that with
decreasing the perpendicular magnetic field the pancake sublattice
transforms from the ``shifted'' sublattice characterized by one
component PV displacement along JVs to the ``trapped'' sublattice
where JVs are occupied by PV rows. The comparison of the free
energies of the tilted lattice and the crossing vortex structure
for the case $\lambda_{ab}>\gamma s$ indicates that in low
($B_x\lesssim\gamma\Phi_0/\lambda_{ab}^2$) and high ($B_x\gtrsim
\Phi_0/\gamma s^2$) in-plane fields, the tilted lattice can exist
if the vector ${\vec B}=B_x{\vec e}_x +B_z{\vec e}_z$ is directed
close to the $c$-axis as well as near the $ab$-plane, while the
crossing lattice is realized in the fields oriented far enough
from the crystal symmetry axes. Furthermore, in the intermediate
in-plane fields the tilted vortex lattice exists only
 at the magnetic field orientations near the $c$-axis whereas
crossing vortex structure settles in the wide angular range until
 the lock-in transition happens.

This paper is organized as follows. The general equations for the
magnetic field distribution and the energy of Josephson vortex in
the presence of the pancake lattice are derived in section
\ref{sec1}. The dense pancake lattice is studied in section
\ref{sec2}. It is shown that, in the limit $\gamma s \gg
\lambda_{ab}$ and $\Phi_0/(\gamma s)^2\ll B_z\ll\Phi_0\gamma^2
s^2/\lambda_{ab}^4$, our model reproduces the results which were
earlier obtained by Koshelev \cite{koshelev}, while the shear
deformation of the PV lattice significantly renormalizes the JV
energy at the higher out-of-plane fields. Section \ref{sec3} is
devoted to the dilute PV lattice. It is described how the novel
vortex substructure with the PV lattice ``trapped'' by the JV
lattice can be realized at low $B_z$. The phase diagram of the
vortex-solid phase in the tilted magnetic fields is considered for
the case of $\lambda_{ab}>\gamma s$ in section \ref{sec4} while
the recent experimental results are discussed in section
\ref{sec5}.

\section{Josephson vortex in the presence of pancake vortex lattice: General equations}
\label{sec1}

We consider a Josephson vortex crossed with the pancake lattice in
the framework of the modified London model. On scales which are
much larger than both the distance between CuO$_2$ planes and the
in-plane coherence length $\xi_{ab}$,  the pancake vortex stack
could be considered as an ordinary vortex line at temperatures
significantly lower than the evaporation temperature \cite{clem,bul-evap}. The
same approach can be also used for the description of the
Josephson vortex far from the nonlinear core. The JV current acts
on PVs through the Lorentz force causing their displacements along
JV, which can be interpreted as a local inclination of the PV
lines away from the $c$-axis. In turn, the local tilt of the PV
stacks induces an additional current along the $c$-axis which
redistributes the ``bare'' JV field. Such physical picture can be
described with one-component PV displacement ${\vec u}=(u,0,0)$
which does not depend on the $x$-coordinate (Fig. 1). The free
energy functional $\F_{PJ}$ can be written as
\begin{eqnarray}
&\F_{PJ}&=\frac{1}{8\pi}\int d^3{\vec R} \Bigl(\hp^2+\rot\hp\Ll\rot\hp
\nonumber \\
&+& \hj^2+ \rot\hj\Ll\rot\hj+ 2\hp\hj+2\rot\hp\Ll\rot\hj\Bigr), \nonumber \\
\label{f0}
\end{eqnarray}
where $\hp$ and $\hj$ are the magnetic fields of PV lines and JV,
respectively, and $\Ll$ is the penetration-depth tensor,
$\nabla=(\partial/\partial x, \partial/\partial
y,\partial/\partial z)$.
In the considered coordinate system the
tensor has only the diagonal components
$\Lambda_{xx}=\Lambda_{yy}=\lambda_{ab}^2$,
$\Lambda_{zz}=\lambda_{c}^2$ with anisotropic penetration depths
$\lambda_{ab}$ and $\lambda_c$. The field $\hp$ is determined by
the displacement $u$ of PVs through the London equation
(see, for instance, \cite{brandt})
\begin{eqnarray}
&\hp&+\rot\left(\Ll\rot\hp\right)= \Phi_0\sum_{i}\int d{\tilde z} \nonumber \\
&\times&\left({\vec e}_z+ \frac{\partial u(Y_i,{\tilde z}){\vec
e}_x}{\partial {\tilde z}}\right) \delta({\vec r} - {\vec
R}_i({\tilde z}) - u(Y_i,{\tilde z}){\vec e}_x), \label{hp0}
\end{eqnarray}
where ${\vec R}_i(\tilde z)=(X_i,Y_i,\tilde z)$ is the equilibrium position
of the $i$-th PV line, ${\vec r}=(x,y,z)$
while ${\vec e}_z$ and ${\vec e}_x$ are unit
vectors along the $z$ and $x$ axes, respectively. (Here we have
accepted that the
parametric equation ${\vec r}_i(\tilde z)$ (with parameter $\tilde z$)
describing the $i$-th vortex line
takes the form of $x(\tilde z)=X_i+u_i(\tilde z)$,
$y(\tilde z)=Y_i$, $z(\tilde z)=z$.) The in-plane
coordinates of the unshifted lines $X_i$ and $Y_i$ are expressed
through the distances $a$ and $b$ (see Fig. 1 b) between PVs and
PV rows as $X_i=al/2+aj$ and $Y_i=bl$ with integer $l$ and $j$. In
our approach, the field of the Josephson vortex also obeys the
London equation
\begin{equation}
h_J-\lambda_{ab}^2\frac{\partial^2h_J}{\partial z^2}-
\lambda_c^2\frac{\partial^2h_J}{\partial
y^2}=\Phi_0\delta(y)\delta(z), \label{hj0}
\end{equation}

\begin{figure}[btp]
\begin{center}\leavevmode
\includegraphics[width=1.15\linewidth]{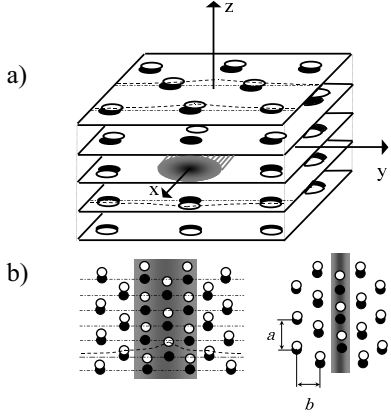}
\caption{The JV crossed by PV stacks which are shifted by the
currents induced by JV: a) 3D sketch depicts the deformation of
the PV lattice in the different CuO$_2$ planes, b) 2D sketches
show the deformation of the PV lattice in a CuO$_2$ plane for the
dense (left) and dilute (right) PV lattices. The filled circles
correspond to the unshifted PVs while the open ones represent the
PVs shifted due to the interaction with JV currents. The shaded
area images the nonlinear JV core region. The dashed-dotted lines
mark the rows of the unshifted PVs while the dashed curve shows
the deviation of PVs from these rows.} \label{f1}
\end{center}
\end{figure}

\noindent where $\delta$-functions should be smoothed on a scale of the
Josephson vortex core. The JV core size along the $z$-axis is
fixed by the interlayer distance $s$, while the core length along
the $y$-direction is limited by the condition that the current
along the $c$-axis can not exceed the maximum interlayer current
$j_c\sim c\Phi_0/(8\pi^2\lambda_c^2s)$. In the presence of PVs,
the current across the layers consists of both the current of JV
itself and the current born by the local inclination of PV lines.
Therefore, the core size along the $y$-axis $\lambda_J$ can be
renormalized in the presence of PVs and should be calculated
self-consistently (see next section). Furthermore, the space
variable $y$ could be replaced by $y-y_0$ in the argument of the
$\delta$-function with $0\leq y_0\leq b$ since the PV lattice can
be arbitrary shifted from the center of JV. However, we take
$y_0=0$ which corresponds to the energetically more
 preferable position \cite{koshelev}.

Next, in order to find the distribution of the magnetic field in
the vortex system and the energy of JV, we will minimize the free
energy (\ref{f0}) as a functional of the displacement $u$. The
fields $\hp$ and $\hj$ can be obtained using  equations
(\ref{hp0},\ref{hj0}) with the displacement $u$ fixed by the
minimization of (\ref{f0}). Then, the  energy of JV, $\E$, defined
as the difference of the free energies (\ref{f0}) with and without
JV, will be derived. This energy includes the self energy of JV
and the change of the free energy of the PV lattice born by the
 interaction with JV. We will use
the elastic approximation, {\it i.e.},
 the free energy (\ref{f0}) and the magnetic field of PVs (\ref{hp0})
 will be expanded up to the second order in $u$.

Using the integral representation of $\delta$-function, the
equation (\ref{hp0}) can be rewritten as
\begin{eqnarray}
&\hp&+\rot\left(\Ll\rot\hp\right) =\Phi_0\sum_i\int d{\tilde
z}\nonumber \\ &\times&\int \frac{d^3{\vec q}}{(2\pi)^3}e^{i{\vec q}{\vec r}}
e^{-i{\vec q}{\vec R}_i(\tilde z)} 
\Biggl({\vec e}_z(1-iq_xu_i(\tilde z)-\frac{1}{2}q_x^2u_i^2(\tilde z))
\nonumber \\ &+&\frac{\partial u_i({\tilde z})}{\partial {\tilde z}}
(1-iq_xu_i(\tilde z)){\vec e}_x\Biggr). 
\label{hp-ap}
\end{eqnarray}
The field $\hp$ of PV lines changes on
different space scales. The first scale is determined by the
characteristic gradient of the displacement $u(y,z)$ and usually is
 much larger than the distance $a$ between PVs. The second
scale is defined by the discreteness of the PV lattice and it is about $a$.
To separate the contribution to the free energy from these scales, we introduce
the Fourier variables
$u(k_y,k_z)$:
\begin{equation}
u(Y_i,z)=\int_{-\pi/b}^{\pi/b}\frac{dk_y}{2\pi}
\int\frac{dk_z}{2\pi}u(k_y,k_z)
e^{i(k_yY_i+k_zz)}, \label{uik}
\end{equation}
and
\begin{equation}
u(k_y,k_z)=b\sum_{Y_i}\int dz u(Y_i,z)e^{-i(k_yY_i+k_zz)},
\end{equation}
where the domain of variation of $k_z$ is restricted by the inequality
$|k_z|\lesssim 1/s$ born by the layerness of the system.
Substituting (\ref{uik}) into (\ref{hp-ap}) and using the
well-known equality $\sum_{i}\int d{\tilde z}\exp{\bigl(i({\vec k}-{\vec
q}){\vec R}_i(\tilde z)\bigr)}=(2\pi)^3(B_z/\Phi_0)\sum_{\vec
Q}\delta({\vec k}-{\vec q}-{\vec Q})$, where ${\vec Q}
=(Q_x,Q_y,0)$ are the vectors of the reciprocal lattice ($Q_x=2\pi
m/a, Q_y=\pi(2n+m)/b$ with integer $m$ and $n$), one gets the
expansion of the field of PVs in series with respect to the
displacement $u$:
\begin{eqnarray}
&&{\vec h}_p=\hp^{(0)}+\hp^{(1)}[u]+\hp^{(2)}[u], \ \ {\vec
n}_p={\vec n}_p^{(0)}+ {\vec n}_p^{(1)}[u]+{\vec n}_p^{(2)}[u],
\nonumber
\\ && \hp^{(i)}+\rot\Ll\rot\hp^{(i)}=
{\vec n}_p^{(i)}
\label{hpi}
\end{eqnarray}
where
\begin{eqnarray}
&{\vec n}&_p^{(0)}={\vec e}_z\Phi_0\sum_{i}\delta_2({\vec
r}^{\perp}-{\vec R}_i^{\perp}),
\nonumber \\
&{\vec n}&_p^{(1)}=B_z\sum_{\vec Q}\int\frac{dk_ydk_z}{4\pi^2}
\nonumber \\ &\times&
({\vec e}_ziQ_x+{\vec e}_xik_z)u(k_y,k_z)
e^{-iQ_xx}e^{i(k_y-Q_y)y}e^{ik_zz},\nonumber \\
&{\vec n}&_p^{(2)}=B_z\sum_{\vec
Q}\int\frac{dk_ydk_zdk_y^{\prime}dk_z^{\prime}}{16\pi^4} u({\vec
k})u({\vec k}^{\prime}) \nonumber \\ &\times&
\left(-\frac{1}{2}Q_x^2{\vec e}_z-Q_xk_z{\vec e}_x\right)
e^{i(k_z+k_z^\prime)z}e^{-iQ_xx}e^{i(k_y+k_y^\prime-Q_y)y}. \nonumber \\
\label{ni}
\end{eqnarray}
The term of $\hp$ with ${\vec Q}=0$ corresponds to the continuous
approximation and varies on the large scale, while terms with ${\vec
Q}\ne 0$ are related to the field components changing on the scale
of $a$.

It is easy to see that only terms with $Q_x=0$ will give
contribution to the part of the free energy describing the
interaction between PVs and JV, because the field $\hj$ does not
depend on $x$ and all terms with $Q_x\ne 0$ vanish after
integration over $x$. Therefore, it is convenient to divide
$\hp^{(1)}$ and ${\vec n}_p^{(1)}$ into two components: ${\vec
n}_p^{(1)}={\vec n}_p^{(Q)}+{\vec e}_x n_p^{*}, \ \ {\vec
h}_p^{(1)}={\vec h}_p^{(Q)}+ {\vec e}_x h_p^{*}$, where
$\hp^{(Q)}$ and ${\vec n}_p^{(Q)}$ include summands with $Q_x\ne
0$ while $h_p^{*}$ and $n_p^{*}$ do not vary with $x$. Then, the
free energy functional $F_{cross}$, containing only terms
dependent on the displacement $u$, can be introduced as
$F_{cross}=\F_{PJ}-\F_P-\F_J$, where $\F_P$ and $\F_J$ are the
free energies of the unperturbed PV lattice and the ``bare'' JV,
respectively. Using equations (\ref{hpi}),
 we obtain the expression for $F_{cross}$ as
\begin{eqnarray}
F_{cross}&=&\frac{1}{8\pi}\int d^3{\vec R}\left({\vec
n}_p^{(Q)}\hp^{(Q)} + 2\hp^{(0)}{\vec
n}_p^{(2)}\right)\nonumber \\ &+&\frac{1}{8\pi}\int d^3{\vec
R}\left(h_p^{*}n_p^{*}+2h_{J}n_p^{*}\right). \label{F-cr-r1}
\end{eqnarray}
The first contribution comes from the terms with $Q_x\ne 0$ and
depends only on the short-scale variations of $\hp$. It is
determined by the shear deformation and the tilt
deformation. The second part describes the interaction of PVs with
JV and with the current generated by PVs along the $y$-axis.
Using the equations (\ref{ni}), the free energy $F_{cross}$
can be rewritten in term of Fourier variables $u(k_y,k_z)$:

\begin{eqnarray}
F_{cross}&=&\frac{1}{2}\int\frac{dk_ydk_z}{4\pi^2}
\left(U_{66}(k_y)+U_{44}(k_y,k_z)\right)u({\vec k})u(-{\vec k}) \nonumber \\
&+&\frac{B_z\Phi_0}{4\pi}\sum_{Q_x=0}\int\frac{dk_ydk_z}{4\pi^2}ik_zu({\vec
k}) \nonumber \\ &\times&\frac{f(k_z,k_y-Q_y)-(B_z/2\Phi_0)ik_zu(-{\vec k})}{1+\lambda_{ab}^2k_z^2+
\lambda_c^2(k_y-Q_y)^2}, 
\label{f-cr-k}
\end{eqnarray}
with the shear energy
\begin{eqnarray}
U_{66}&=&\frac{B_z^2}{4\pi}\sum_{Q_x\ne 0}\Biggl\{
\frac{Q_x^2}{1+\lambda_{ab}^2Q_x^2+\lambda_{ab}^2(k_y-Q_y)^2)} \nonumber \\
&-&
\frac{Q_x^2}{1+\lambda_{ab}^2(Q_x^2+Q_y^2)}\Biggr\},
\label{shear0}
\end{eqnarray}
and the tilt energy
\begin{eqnarray}
U_{44}&=&\frac{B_z^2}{4\pi}\sum_{Q_x\ne 0}\Biggl\{
\frac{Q_x^2}{1+\lambda_{ab}^2k_z^2+\lambda_{ab}^2Q_x^2+\lambda_{ab}^2(k_y-Q_y)^2}
\nonumber\\&-&\frac{Q_x^2}{1+\lambda_{ab}^2Q_x^2+\lambda_{ab}^2(Q_y-k_y)^2} \nonumber \\
&+&\frac{k_z^2}{1+\lambda_{ab}^2k_z^2+\lambda_c^2Q_x^2+\lambda_c^2(k_y-Q_y)^2}
\nonumber \\ &+& \frac{(\lambda_c^2-\lambda_{ab}^2)Q_x^2k_z^2}{
\left(1+\lambda_{ab}^2k_z^2+\lambda_{ab}^2Q_x^2+\lambda_{ab}^2(k_y-Q_y)^2\right)} \nonumber \\
&\times&\frac{1}{
\left(1+\lambda_c^2Q_x^2+\lambda_c^2(k_y-Q_y)^2+\lambda_{ab}^2k_z^2\right)}\Biggr\}.
\end{eqnarray}
The expressions for $U_{44}$ and $U_{66}$ represent sums over the
reciprocal lattice vectors with $Q_x\ne 0$, while the summation in
the last term of (\ref{f-cr-k}) is performed only over the
reciprocal lattice vectors with $Q_x=0$. The function $f({\vec
q})$ in (\ref{f-cr-k}) appears due to smoothing of $\delta$
function in eq.~(\ref{hj0}) and can be evaluated as $f({\vec
q})\approx 1$ in the rectangular region $|q_z|\leq 1/s$,
$|q_y|\leq 1/\lambda_J$ and $f\approx 0$ outside that area. The
summation in the expression (\ref{shear0}) for the shear elastic
energy was done by Brandt \cite{bran} in the limit $k_y\ll \pi/b$:
\begin{equation}
U_{66}=C_{66} k_y^2,
\label{U660}
\end{equation}
where the shear elastic modulus $C_{66}$ is expressed as
$C_{66}=(B_z\Phi_0)/(8\pi\lambda_{ab})^2$ for
$a_0=\sqrt{\Phi_0/B_z}<\lambda_{ab}$, while
$C_{66}=\sqrt{\pi\lambda_{ab}/(6a_0)}\Phi_0^2/(4\pi\lambda_{ab}^2)^2\exp(-a_0/\lambda_{ab})$
for $a_0>\lambda_{ab}$. The tilt energy was obtained in \cite{brandt,kosh-kes}:
\begin{eqnarray}
U_{44}&=&\frac{B_z\Phi_0}{32\pi^2\lambda_{ab}^4}\Biggl(\ln\left(1+\frac{k_z^2}{\lambda_{ab}^{-2}
+K_0^2}\right) \nonumber \\&+&\frac{k_z^2\lambda_{ab}^4}{\lambda_c^2}
\ln\left(\frac{\xi_{ab}^{-2}}{K_0^2+(k_z^2/\gamma^2)+\lambda_c^{-2}}\right)
\Biggr) \label{u44-br}
\end{eqnarray}
for $k_z \gtrsim K_0=2\pi/b$, while
\begin{equation}
U_{44}=\left(3.68\frac{\Phi_0^2}{(4\pi\lambda_{ab})^4}+\frac{B_z\Phi_0\ln(a^2/\xi_{ab}^2)}
{32\pi^2\lambda_c^2}\right)k_z^2=C_{44}^{eff}k_z^2 \label{u441}
\end{equation}
for $k_z\ll K_0$. Performing summation over $Q_y$ in the second
term of equation (\ref{f-cr-k}) (see Appendix A),
we finally obtain the free energy functional:
\begin{eqnarray}
&&F_{cross}=\int\frac{dk_y}{2\pi}\frac{dk_z}{2\pi}
\Biggl\{\frac{1}{2}\left(U_{44}+U_{66}\right)u({\vec k})u(-{\vec
k})
\nonumber \\
&&+\frac{B_z}{4\pi} ik_z\Psi(k_z,k_y)\left(u({\vec
k})-ik_z\frac{B_z}{2\Phi_0}u({\vec k})u(-{\vec k})\right)\Biggr\},
\label{f-cr-last}
\end{eqnarray}
where $\Psi$ is defined by the equation
\begin{eqnarray}
\Psi(k_z,
k_y)&=&\frac{\Phi_0b}{2\lambda_c\sqrt{1+\lambda_{ab}^2k_z^2}} \nonumber \\
&\times&
\frac{\sh\left(\sqrt{1+\lambda_{ab}^2k_z^2}b/\lambda_c\right)}{
\ch\left(\sqrt{1+\lambda_{ab}^2k_z^2}b/\lambda_c\right)-\cos k_yb}
\label{psi0}
\end{eqnarray}
for $k_y<\min({\pi/b,\ 1/\lambda_J})$ and $k_z< 1/s$ while
$\Psi\approx 0$ outside that rectangular area. In the case of
small values of wave vector ${\vec k}$ ($k_y\ll \pi/b$ and $k_z\ll
\gamma/b$), the discreteness of PV lattice is irrelevant and the
function $\Psi$ coincides with the Fourrier image of the ``bare''
JV field, but $\Psi$ is modified substantially for larger $k_y$ or
$k_z$.

The minimization of the free energy functional (\ref{f-cr-last})
determines the displacement $u$ as
\begin{equation}
u({\vec k})=\frac{B_z}{4\pi} \frac{ik_z\Psi({\vec
k})}{U_{44}+U_{66}+(B_z^2k_z^2/4\pi\Phi_0)\Psi({\vec k})}.
\label{u}
\end{equation}
In order to describe the field distribution of a JV in the
crossing lattice, the averaged magnetic induction $\bj$ along the
$x$-direction generated by both JV and inclined PV lines can be
introduced. By substituting the found displacement (\ref{u}) into
equations (\ref{hpi},\ref{ni}), the magnetic induction of JV
$\bj=h_J+h_p^*$  is rewritten as
\begin{eqnarray}
&&\bj=\int\frac{dq_zdq_y}{(2\pi)^2}\frac{\Phi_0e^{iq_yy+iq_zz}}
{1+\lambda_c^2q_y^2+\lambda_{ab}^2q_z^2}
\nonumber \\
&&+B_z\sum_{Q_y}\int\frac{dk_ydk_z}{(2\pi)^2} \frac{ik_zu({\vec
k})}{1+\lambda_c^2(k_y-Q_y)^2+k_z^2\lambda_{ab}^2}e^{i(k_y-Q_y)+ik_zz},
\nonumber \\
\label{b0}
\end{eqnarray}
where $q_y$ and $q_z$ are the wave vectors of a ``bare'' JV
($|q_y|<1/\lambda_J$, $|q_z| < 1/s$), while the wave vectors
${\vec k}$ of the PV lattice are restricted also by the first
Brillouin zone of the PV lattice ($|k_y|<\min({1/\lambda_J,\
\pi/b})$, $|k_z|<1/s$). To get the  energy $\E$ it is necessary to
add $F_{cross}$ to the energy of the JV itself. Obviously, the
energy of a JV in the presence of PV lines is always lower than
one of a ``bare'' JV. Indeed, for the displacement of PVs $u$
determined by equation (\ref{u}), the energy $F_{cross}$ takes the
minimum value which is smaller than zero, since $F_{cross}=0$ at
$u=0$. Finally, the energy of JV in the crossing lattice obeys the
equation
\begin{eqnarray}
&&\E=\frac{\Phi_0^2}{8\pi}\int \frac{dq_ydq_z}{(2\pi)^2}\frac{1}{1+\lambda_c^2q_y^2+\lambda_{ab}^2q_z^2}
\nonumber \\
&&-\frac{B_z^2}{32\pi^2}\int\frac{dk_ydk_z}{(2\pi)^2}
\frac{k_z^2\Psi({\vec k})\Psi(-{\vec
k})}{U_{44}+U_{66}+(k_z^2B_z^2/4\pi \Phi_0)\Psi({\vec k})}.
\label{e0}
\end{eqnarray}

Equations (\ref{u}-\ref{e0}) together with the condition that the
current density along the $c$-axis should be smaller than the
maximum current density $j_c$, determine completely the behavior
of the PV lines and the JV. However, in further analysis it is
convenient to investigate the dense ($\gamma s \gg a$) and dilute
($\gamma s\ll a$) pancake lattices separately.

\section{Josephson vortex in the presence of dense PV lattice}
\label{sec2}

For the case of the dense PV lattice, many PV rows are placed on
the nonlinear JV core (Fig. 1b, left sketch). It means that the
magnetic field of a bare JV varies on scales larger than the
distance between PV lines even near the JV core. Thus, the
continuous approximation is applicable in the whole space. In this
case $|k_y|<1/\lambda_J<\pi/b$ and, therefore, the cosine and
hyperbolic functions in (\ref{psi0}) can be expanded in the
series. Hence, the function $\Psi$ can be rewritten as
\begin{equation}
\Psi=\frac{\Phi_0}{1+\lambda_{ab}^2k_z^2+\lambda_c^2k_y^2}.
\label{psi-con}
\end{equation}
Substituting this expression for $\Psi$ into (\ref{b0},\ref{e0})
and omitting the difference between ${\vec k}$ and ${\vec q}$, the
equations  for the dense PV lattice (which determine the field
distribution and the energy of JV) are deduced as

\begin{eqnarray}
&&\bj=\int^{1/\lambda_J}_{-1/\lambda_J}\frac{dq_y}{2\pi}\int_{-1/s}^{1/s}
\frac{dq_z}{2\pi} \nonumber \\ &\times&
\frac{\Phi_0e^{iq_yy+iq_zz}}{1+\lambda_c^2q_y^2+
\lambda_{ab}^2q_z^2+q_z^2B_z^2/(4\pi(U_{44}+C_{66}k_y^2))},
\end{eqnarray}
and
\begin{eqnarray}
&&\E=\frac{\Phi_0^2}{8\pi}\int_{-1/\lambda_J}^{1/\lambda_J}
\frac{dq_y}{2\pi}\int_{-1/s}^{1/s}\frac{dq_z}{2\pi} \nonumber\\ &\times&
\frac{1}{1+\lambda_c^2q_y^2
+\lambda_{ab}^2q_z^2+B_z^2q_z^2/(4\pi(U_{44}+C_{66}q_y^2))}.
\label{ej-a<lambdaJ}
\end{eqnarray}
The last undefined parameter, $\lambda_{J}$, can be obtained from
the condition $|\partial \bj(y\approx\lambda_J, z=0)/\partial y|
\sim(4\pi/c)j_c$:
\begin{eqnarray}
&&\frac{\pi}{\lambda_c^2
s}\simeq\lambda_J\int_{-1/\lambda_J}^{1/\lambda_J}dq_y
\int_{-1/s}^{1/s}dq_z \nonumber \\&\times& \frac{q_y^2}{1+\lambda_c^2q_y^2+
\lambda_{ab}^2q_z^2+q_z^2B_z^2/(4\pi(U_{44}+C_{66}q_y^2))}.
\label{lambdaJ}
\end{eqnarray}

The region of integration is shown in Fig. 2a. The rectangular
domain of possible wave vectors replaces the usual elliptical one
due to a peculiar core structure of JV. In anisotropic London
model, the core of an ordinary vortex is defined by the elliptical
stream line of the persistent current having the depairing value
\cite{kog}. However, in our case the maximum value of $q_z$ is
determined by the layerness of the medium while the largest value
of $q_y$ is restricted by the Josephson critical current along the
$c$-axis. The rectangular domain of wave vectors (Fig. 2 a) can be
divided into
 ``screened'' ($|k_z|\gtrsim 1/b$) and ``remote''($|k_z|<1/b$) subdomains.
The first one corresponds to the
region where one can roughly neglect the weak logarithmical dependence
 on $k_z$ in eq. (\ref{u44-br}) to express the tilt energy as:
\begin{equation}
U_{44}\approx \avu+\avc k_z^2,
 \label{uav}
\end{equation}
with $\avu=(B_z\Phi_0/32\pi^2\lambda_{ab}^4)
\ln(1+\overline{k_z}^2/(\lambda_{ab}^{-2}+b^{-2}))$,
$\avc=\frac{B_z\Phi_0}{32\pi^2\lambda_{c}^2}
\ln\left(\frac{\xi_{ab}^{-2}}{b^{-2}+(\overline{k_z}^2/\gamma^2)+\lambda_c^{-2}}\right)$
and $\overline{k_z}\sim\sqrt{1/bs}$. 
\begin{figure}[btp]
\begin{center}\leavevmode
\includegraphics[width=1.15\linewidth]{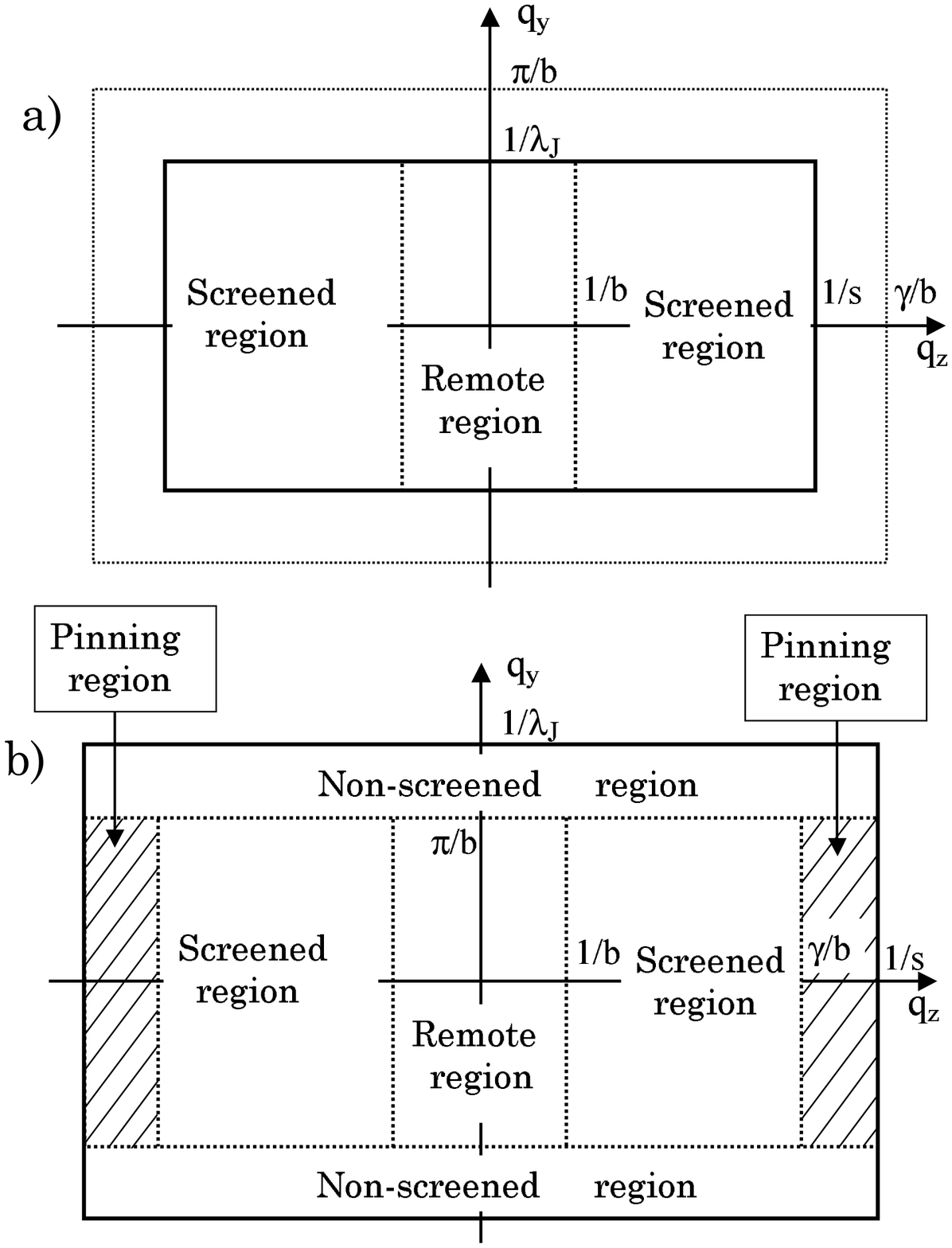}
\caption{The integration region in equations (\ref{b0}) and
(\ref{e0}) for the dense PV lattice (a) and the dilute PV lattice
(b). In the dense case (a), the region of the available wave
vectors ${\vec k}$ of the PV lattice coincides with the accessible
domain of wave vectors ${\vec q}$;
 the continuous approximation (\ref{psi-con}) for $\Psi$ is always valid.
 The approximation (\ref{uav}) is correct in the ``screened'' domain of
wave vectors, while it fails in the ``remote'' region. In the
dilute case (b), the continuous approximation is correct only in
the region $|q_y|\ll\pi/b;\ |q_z|\ll\gamma/b$; the
``non-screened'' region $|q_y|>\pi/b$ is not accessible for ${\vec
k}$. The continuous approximation for $\Psi({\vec k})$ is broken
in the ``pinning'' region $|k_y|<\pi/b;\ \gamma/b\lesssim |k_z|\lesssim 1/s$.}
\label{f2}
\end{center}
\end{figure}

\noindent The expression
(\ref{uav}) is completely wrong in the ``remote'' region in which it is
necessary to use equation (\ref{u441}).
By using approximation (\ref{uav}) for the tilt energy,
the integrals in equations (\ref{ej-a<lambdaJ}) and (\ref{lambdaJ})
are easily evaluated (Appendix B) and we get
\begin{equation}
\lambda_{J}\approx\max{\left(\frac{\lambda_cs}{\lambda_{ab}^{eff}},\sqrt{\frac{\lambda_cs}{
\lambda_{ab}^{eff}}\lambda_{ab}}\right)}
\label{lambdaJ-K}
\end{equation}
and
\begin{equation}
\E\approx\frac{\Phi_0^2}{16\pi^2\lambda_{ab}^{eff}\lambda_c}\left(\sqrt{\frac{C_{66}}
{\avu\lambda_{ab}^2}}
\frac{\lambda_{ab}}{\lambda_{J}}+\ln\frac{\lambda_{cut}}{\lambda_J}\right),
\label{ej-k}
\end{equation}
where $\lambda_{cut}\sim \min(\lambda_{c}, \min(\gamma b,
\max(b\lambda_c/\lambda_{ab}^{eff},
\sqrt{b\lambda_c\lambda_{ab}/\lambda_{ab}^{eff}})))$, while the
renormalized penetration depth $\lambda_{ab}^{eff}$ is expressed
as
\begin{equation}
\lambda_{ab}^{eff}=\sqrt{\lambda_{ab}^2+\frac{B_z^2}{4\pi\avu}}.
\label{lam-eff}
\end{equation}
The physical reason of the renormalization of the in-plane penetration
depth is related to the screening of the JV field by currents born by the local
 inclination of PV lines. From eq.~(\ref{lambdaJ-K}) it is easy to see
that the size of the nonlinear JV core also decreases due to the
interaction of JV and PVs. The similar conclusion was given
earlier by Koshelev \cite{koshelev}, who considered the additional
phase variation of the order parameter born by the displacement of
PVs. However, the shear contribution to the renormalization of
$\E$ and $\lambda_J$ was neglected in \cite{koshelev}, which could
be done only for $\lambda_J>\lambda_{ab}$ (see equations
(\ref{lambdaJ-K}) and (\ref{ej-k})). In the opposite case, {\it
i.e.}, when the London penetration depth exceeds the JV core size,
the shear deformation becomes relevant and, as a result,
$\lambda_J$ decreases with $B_z$ slower than it was proposed in
\cite{koshelev}.

To understand how the field of JV is distributed in
 the real space, we
rederive the results considering the free energy functional of the
displacement $u$ defined as a function of the spatial coordinates.
In the limit $\gamma s\gg a$, the JV field varies on scales larger
than the distance between PVs even near the JV core. This means
that the field $\hp$ along the $x$-axis can be averaged out on the
scale larger than $a$:
\begin{equation}
h_p^*-\lambda_{ab}^2\frac{\partial^2h_p^*}{\partial z^2}-
\lambda_c^2\frac{\partial^2h_p^*}{\partial y^2}=n^*=B_z
\frac{\partial u}{\partial z}.
\label{hp*-r}
\end{equation}
However, the short range variations of the field $\hp$
 give the shear energy $U_{66}$ and the tilt energy $U_{44}$.
After ignoring the slow logarithmic dependence on ${\vec k}$ in
the expression for $U_{44}$, one can conclude that the density of
the tilt energy in the real space is $\avu u^2(y,z)$, while the
density of the shear energy is
 $U_{66}=C_{66}(\partial u/\partial
y)^2$. Thus, the free energy functional is expressed as
\begin{eqnarray}
F_{cross}&=&\frac{1}{8\pi}\int d^3{\vec R} \Biggl(4\pi
C_{66}\left(\frac{\partial u}{\partial y}\right)^2 \nonumber \\ &+& 4\pi \avu u^2
+ \hp^{*}B_z\frac{\partial{\vec u}}{\partial z}+
2\hj\frac{\partial{\vec u}}{\partial z}\Biggr). \label{free+el}
\end{eqnarray}
The first three terms represent the elastic energy (born by shear,
electromagnetic tilt and Josephson coupling tilt rigidity, respectively),
but the last
term is related to the interaction of the PV lines with the
current generated by JV.

In order to get the complete set of equations for the displacement
$u$ and the averaged magnetic induction $\bj$, we have minimized
the functional (\ref{free+el}) and have added together equations
(\ref{hj0}) and (\ref{hp*-r}):
\begin{eqnarray}
&&-4\pi C_{66}\frac{\partial^2 u}{\partial y^2}+4\pi \avu u -
2B_z\frac{\partial \bj}{\partial z}=0,
\nonumber\\
&&\bj-\lambda_{ab}^2\frac{\partial^2\bj}{\partial z^2}-
\lambda_c^2\frac{\partial^2\bj}{\partial y^2}=\Phi_0\delta(y)\delta(z)+
B_z\frac{\partial u}{\partial z}.
\label{set}
\end{eqnarray}
This set of equations is applicable if the continuous
approximation is valid ($\lambda_J \gg a$) and, strictly speaking,
only when the tilt energy $U_{44}(k_z)$ can be replaced by the
constant $\avu$. The last condition fails on the distances far
from JV ($z^2+y^2/\gamma^2>b^2$). In this ``remote'' region, the
constant $\avu$ has to be substituted by
 $-C_{44}^{eff}\partial^2/\partial z^2$. Besides, if
$\lambda_{ab}>\gamma s$, the parameter $\avu$ should be replaced
by $-\avc \partial^2/\partial z^2$ near the JV core
($z<\lambda_{ab}/\gamma$).

Even though we consider only the situation when the set of
equation (\ref{set}) is valid, {\it i.e.}, the case
$\lambda_{ab}<\gamma s$ and the region $z^2+y^2/\gamma^2<b^2$, the
solution of equations (\ref{set}) seems to be quite complicated.
The relation between the displacement $u$ and the magnetic
induction $\bj$, which is obtained from the first equation of
(\ref{set}), becomes nonlocal due to the shear rigidity of the PV
lattice:
\begin{equation}
u=\frac{B_z}{8\pi\sqrt{C_{66}\avu}}\int_{-\infty}^{\infty}d{\tilde y}
\frac{\partial \bj({\tilde y}, z)}{\partial z} \ e^{-|y-{\tilde y}|/\delta}
\label{ub}
\end{equation}
where $\delta=\lambda_{ab}\sqrt{C_{66}/(\avu\lambda_{ab}^2)}\sim
\lambda_{ab}$ is the characteristic length of a nonlocality.
However, the nonlocality is irrelevant if the space scale of the
variation of $\bj$ is substantially large then $\delta$, {\it
i.e.}, if $\lambda_J\gg \lambda_{ab}$. In such a case, the
equations (\ref{set}) for $\bj$ and $u$ can be decoupled
\begin{eqnarray}
&&u=\frac{B_z}{4\pi U_{44}}\frac{\partial \bj}{\partial z},
\nonumber
\\ &&\bj-(\lambda_{ab}^{eff})^2\frac{\partial^2 \bj}{\partial
z^2}- \lambda_c^2\frac{\partial^2 \bj}{\partial
y^2}=\Phi_0\delta(y)\delta(z).
\label{bjj-1}
\end{eqnarray}
The equation (\ref{bjj-1}) for induction $\bj$ is the London equation with the
 renormalized in-plane penetration depth $\lambda_{ab}^{eff}$.
Therefore, the field distribution $\bj$, not far from the center
of the Josephson vortex ($z^2+y^2/\gamma^2\lesssim b^2$), can be
approximated as
\begin{equation}
\bj=\frac{\Phi_0}{2\pi\lambda^{eff}_{ab}\lambda_c}
{\rm K_0}\left(\sqrt{z^2/(\lambda^{eff}_{ab})^2+
y^2/\lambda_c^2}\right),
\label{bj-real}
\end{equation}
where ${\rm K_0}(x)$ is a modified Bessel function of zero order.
Using the free energy functional (\ref{free+el}) and equations
(\ref{set}), it is easy to show that the energy of JV is
determined by the field in its center, {\it i.e.},
$\E=\Phi_0/(8\pi)\bj(y\approx \lambda_J,z\approx s)$:
\begin{equation}
\E=\frac{\Phi_0^2}{16\pi^2\lambda^{eff}_{ab}\lambda_c}\ln(\lambda_{ab}^{*}/s)
\label{ej-real}
\end{equation}
with the length $\lambda_{ab}^{*}=\lambda_{ab}^{eff}$. However,
the set of equations (\ref{set}) becomes incorrect in the region
$z^2+y^2/\gamma^2>b^2$ which cuts off that length as
$\lambda_{ab}^{*}\approx b$. Thus, the expression (\ref{ej-real})
coincides with the earlier obtained equation (\ref{ej-k}) in the
studied case $\lambda_J>\lambda_{ab}$. The results
(\ref{bj-real},\ref{ej-real}) can be interpreted in terms of the
effective anisotropy parameter
$\gamma^{eff}=\lambda_c/\lambda_{ab}^{eff}$ which governs the JV
lattice. Since $\lambda_{ab}^{eff}>\lambda_{ab}$, the effective
anisotropy $\gamma^{eff}$ is reduced in the presence of PVs with
respect to the ``bare'' one $\gamma=\lambda_c/\lambda_{ab}$. The
similar anisotropy $\gamma^{eff}$ was earlier introduced
 \cite{koshelev} as a ratio $\gamma^{eff}=\lambda_J/s$, but these
two different definitions of $\gamma^{eff}$ give the same value
 in the case $\lambda_J>\lambda_{ab}$ when the
shear deformation is irrelevant.

Here, we discuss how the core size and the JV energy are changed
with the magnetic induction $B_z$ if $\gamma s>\lambda_{ab}$. For
quite high magnetic inductions $B_z\gtrsim
B_1=(\Phi_0/\lambda_{ab}^2)\times(\gamma s/\lambda_{ab})^2$, the
size of the nonlinear core $\lambda_J$ is smaller than
$\lambda_{ab}$ and the shear contribution to the free energy is
important. The second logarithmic term in the JV energy
(\ref{ej-k}) can be omitted, and the core size obeys the equation
$\lambda_J(B_z)\simeq\sqrt{\gamma s a}$.
 With decreasing of induction, the core size increases
proportionally to $B_z^{-1/4}$ and reaches $\lambda_{ab}$ at
$B_z\approx B_1$. At low fields, the shear interaction between
rows is irrelevant, the JV core size becomes
$\lambda_{J}=\lambda_cs/\lambda_{ab}^{eff}\approx \gamma s
a/\lambda_{ab}$, and the energy of JV is determined by the
logarithmic term in (\ref{ej-k}). Below the field
$\Phi_0/\lambda_{ab}^2$, at which the distance $a$ between PVs
exceeds $\lambda_{ab}$, the currents generated by PVs practically
do not influence on the JV field and, thus, the renormalization of
$\lambda_{ab}$, $\lambda_{J}$ and the $\E$ vanishes. In the case
of $\lambda_{ab}>\gamma s$ the physical picture is different from
the previous situation. The core size obeys the law
$\lambda_J\simeq\sqrt{\gamma s a}$ at fields $B_z>\Phi_0/(\gamma
s)^2$. Below this field, the effective value of the in-plane
London penetration depth $\lambda_{ab}^{eff}\sim \lambda_{ab}^2/a$
(\ref{lam-eff}) is
 still larger than $\lambda_{ab}$, while the JV core size is saturated
 as $\lambda_J=\gamma s$. This means that JV field shows different
behavior far from JV ($z^2+y^2/\gamma^2>b^2/\gamma^2$), where the
redistribution due to the local inclination of PV lines is still
important, and close to the JV core.

\section{Josephson vortex in the presence of dilute PV lattice.}
\label{sec3}

Far from the JV center, $z^2+y^2/\gamma^2>b^2/\gamma^2$, the JV
field varies slowly which causes the smooth variation of the
displacement $u$ even for the case of the dilute PV lattice
($a>\gamma s$). In that spatial region, the continuous
approximation is still valid. On the other hand, near the JV core
($|y|<b$), the JV current increases quite fast inducing a large
displacement of the PV stack placed on the center of JV. In this
case, the continuous approximation is not applicable. To describe
such physical situation, we consider the wave vector area of
${\vec k}$ divided into two domains (Fig. 2b). In the first
interval $|k_y|<\pi/b\ {\rm and} \ |k_z|<\gamma/b$, the function
$\Psi$ can be still roughly approximated by equation
$\Psi\approx\Phi_0/(1+\lambda_c^2k_y^2+\lambda_{ab}^2k_z^2)$,
while $\Psi\approx\Phi_0b/(2\lambda_c\lambda_{ab}k_z)$ in the
second region $|k_y|<\pi/b\ {\rm and} \ \gamma/b<|k_z|<1/s$
(``pinning'' region in Fig. 2b). Following this approach, the
energy of JV is evaluated as
\begin{eqnarray}
&&\E(a\gg \gamma s)
\approx\frac{\Phi_0^2}{8\pi}\int_{-\pi/b}^{\pi/b}
\frac{dq_y}{2\pi}\int_{-\gamma/b}^{\gamma/b}\frac{dq_z}{2\pi} \nonumber \\
&\times&\frac{1}{1+\lambda_c^2q_y^2
+\lambda_{ab}^2q_z^2+B_z^2q_z^2/(4\pi(U_{44}+U_{66}))}\nonumber\\
&+&\frac{\Phi_0^2}{16\pi^2\lambda_c\lambda_{ab}}\ln\left(\frac{b}{\gamma
s}\right) \nonumber \\
&-&\frac{B_z\Phi_0^3}{128\pi^3a\lambda_c^2\lambda_{ab}^2}\int_{\gamma/b}^{1/s}
\frac{dk_z}{U_{44}+B_z\Phi_0k_z/(8\pi\lambda_c\lambda_{ab}a)}.
\label{e+pin}
\end{eqnarray}
The first term comes from the spatial region far from the center
of JV while the second and the third terms are related to the
vicinity of the JV center. The screening of the ``bare'' JV field
vanishes near JV (``non-screened'' region in ${\vec k}$-space)
which
 determines the second term in (\ref{e+pin}). The last term in
(\ref{e+pin}) represents the energy gain due to the strong
interaction between the PV line placed on the JV core and the JV
currents (the energy gain of a PV stack placed on a JV in the
limit $\gamma s\gg\lambda_{ab}$ and $B_z\rightarrow 0$ was
calculated by Koshelev \cite{koshelev}). Since the last term is
sensitive to the mutual position of JV and the nearest PV line,
this contribution can be called as the ``crossing lattice
pinning''. Using the results of Appendix B and taking into account
that the evaluation
$B_z\Phi_0k_z/(8\lambda_c\lambda_{ab}a)\lesssim\avc k_z^2$ is held
in the ``pinning region'' ($k_z>\gamma/b$), the energy of JV is
finally obtained:
\begin{eqnarray}
&&\E\approx\frac{\Phi_0^2}{16\pi^2\lambda_c\lambda_{ab}^{eff}}\left(
\frac{2\mu_1\avc\gamma^2}{\pi\avu\lambda_{ab}^2}\frac{\lambda_{ab}^3}
{b^2\lambda_{ab}^{eff}}+
\sqrt{\frac{\mu_2^2C_{66}}{\avu\lambda_{ab}^2}}\frac{\lambda_{ab}}{b}\right) \nonumber \\
&+&\frac{\Phi_0^2}{16\pi^2\lambda_c\lambda_{ab}^{eff}}
\ln\left(\frac{\lambda_{cut}}{b}\right) +
\frac{\phi_0^2}{16\pi^2\lambda_c\lambda_{ab}}
\ln\left(\frac{b}{\gamma s}\right) 
\nonumber \\ &-& \mu\frac{\Phi_0^2}{4\pi
a\lambda_c} \arctan \left(\frac{b-\gamma
 s}{\sqrt{\avu/\avc}sb+\gamma\sqrt{\avc/\avu}}\right),
\label{e+pin-g}
\end{eqnarray}
where $\mu=B_z\Phi_0/(32\pi^2\lambda_c\lambda_{ab}^2\sqrt{\avc\avu})<1$
is the dimensionless function depending quite slowly on $B_z$ and
the numerical parameters $\mu_1$ and $\mu_2$ are about unity.

Next, we will discuss how the renormalization of the JV energy comes
in with increasing of the $z$-component of the magnetic field. At
low fields, $B_z\ll \Phi_0/\lambda_{ab}^2$ ($a\gg\lambda_{ab}$),
the first term and the last term in (\ref{e+pin-g}) can be omitted
and the expression for the energy of a ``bare'' JV
reported earlier in \cite{clem1,koshelev1} is reproduced
\begin{equation}
\E=\frac{\Phi_0^2}{16\pi^2\lambda_c\lambda_{ab}}\ln\left(\frac{\lambda_{c}}{\gamma
s} \right). \label{ebare}
\end{equation}
For the case $\lambda_{ab}>\gamma s$, the renormalization of JV
energy becomes relevant at $B_z\approx\Phi_0/\lambda_{ab}^2$, {\it
i.e.}, earlier than the JV core size starts to decrease which
occurs only in fields $B_z>\Phi_0/(\gamma s)^2$. The origin of
this behavior is that the additional current along the $c$-axis
induced by tilted PV stacks is much smaller than $j_c$ near JV
core in the field interval $\Phi_0/\lambda_{ab}^2\lesssim
B_z\lesssim\Phi_0/(\gamma s)^2$, but the inclination of all PV
lines can still cause the renormalization of the JV field on
scales larger than $a$.
In the field interval $\Phi_0/\lambda_{ab}^2< B_z
\lesssim\Phi_0/\gamma^2 s^2$, the main contribution to the Josephson
vortex energy (\ref{e+pin-g}) comes from the first term related to
the tilt elastic
rigidity (born by Josephson coupling of PVs) and shear elasticity
of the PV lattice. Strictly speaking, from our rough
estimation of (\ref{e0}), we can not conclude how strongly $\E$ is
suppressed in that field interval, {\it i.e.} in the presence of
the dilute PV lattice. Nevertheless, the pinning energy (last term
in (\ref{e+pin-g})) could be the same order of magnitude as the
first and the second terms in (\ref{e+pin-g}) in fields
$B_z\sim \Phi_0/\lambda_{ab}^2$ and may decrease $\E$ substantially.

Another interesting possibility arising due to the ``crossing
lattice pinning'' is the rearrangement of the PV lattice in the
presence of the JV sublattice. In the in-plane magnetic fields
$B_x$, JVs form a triangular lattice with distances $a_J$ and
$b_J$ between JVs (see inset in Fig. 3a). In general, the PV
sublattice and the JV sublattice are not commensurate: $a_J\ne pb$
with integer $p$. This means that the considered one-component
 displacement of PVs ${\vec u}=(u(y,z),0,0)$
(the ``shifted''  PV lattice shown in Fig. 3a) does not provide
the
 energy gain coming from
 the ``crossing lattice pinning'' since the PV rows can not occupy the centers
of JVs. However, the PVs can be rearranged in order to occupy all
JVs (the ``trapped'' PV lattice shown in Fig. 3b) if the PV lines
shift also along the $y$-direction: ${\vec u} =
(u_x(x,y,z),u_y(x,y,z),0)$. The ``crossing lattice pinning''
decreases the free energy of the ``trapped'' PV lattice, while the
additional shear deformation
 acts in the opposite way through increasing
the free energy. For the case $B_x<\gamma B_z$, the energy gain
related to the ``trapped'' PV lattice is calculated by normalizing
the last term of equation (\ref{e+pin-g}) per unit volume:
\begin{equation}
E_{tr}=\mu\frac{B_x\Phi_0}{4\pi a\lambda_c}\arctan\left(
\frac{b-\gamma s}{\sqrt{\avu/\avc}sb+\gamma\sqrt{\avc/\avu}}\right).
\label{trap}
\end{equation}
But, in order to trap the PV lattice, the total displacement of
PVs along the $y$-axis between the two nearest JV rows, {\it
i.e.}, on the scale $a_J$, should be about $b$. Following the
simple analysis \cite{rakh}, the extra shear deformation (inset in
Fig. 3b) is about $\delta b/b\sim b/a_J$ ($\delta b$ is the change
of the distance between rows of PVs) and the energy loss
$E_{shear}$ can be estimated as:
\begin{equation}
E_{shear}\simeq\nu C_{66}\left(\frac{b}{a_J}\right)^2\approx
\nu C_{66}\frac{B_x}{\gamma B_z}
\label{shear}
\end{equation}
with numerical constant $\nu\lesssim 1$. For the case $\gamma
s\gg\lambda_{ab}$, the shear elastic energy (\ref{shear}) is
strongly suppressed in the fields $B_z<\Phi_0/(\gamma s)^2$ where
the ``crossing lattice pinning'' is active, since $C_{66}$ is
exponentially small if $a>\lambda_{ab}$ (\ref{U660}).

\begin{figure}[btp]
\begin{center}\leavevmode
\includegraphics[width=1.15\linewidth]{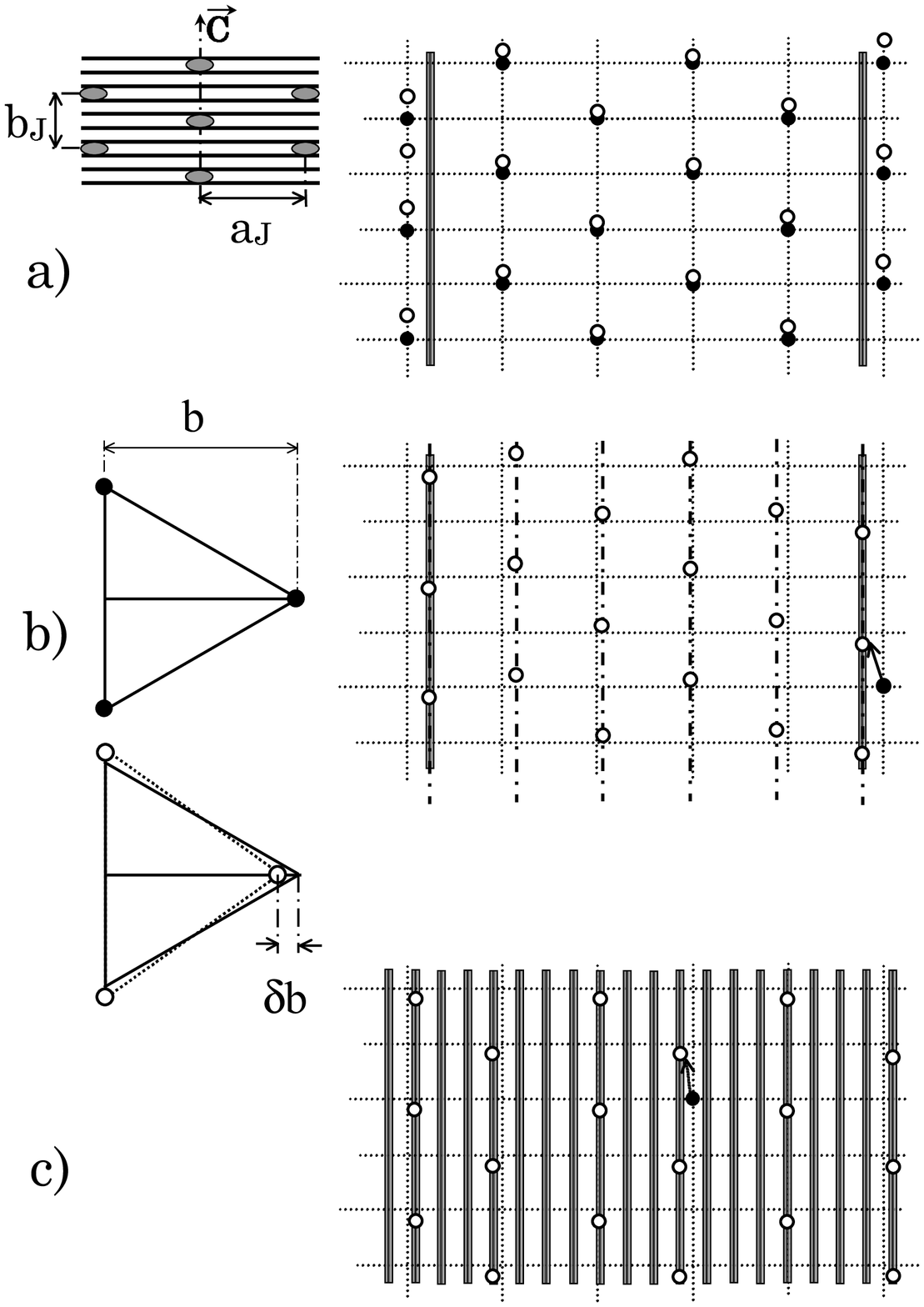}
\caption{The different substructures of the crossing lattice: a)
the ``shifted'' PV lattice characterized by one-component
displacement along JVs, b) the PV lattice trapped by the JV
sublattice for the case when the distance $a_J$ between JVs
exceeds the distance $b$ between PV rows, c) the ``trapped'' PV
lattice for $a_J<b$, {\it i.e.}, in the case of the field
orientations
 very close to the $ab$-plane. The dotted lines depict the rows of
 unperturbed lattice (crossings of these lines and filled circles mark the
 positions of unshifted PVs) in all sketches. Dashed-dotted lines
 indicate the rows of the ``trapped'' PV lattice which are deformed
 in order to match with JV sublattice. The arrows directed from the filled
 circles to the open ones show the two component displacement of PVs
 for the cases represented in sketches b) and c).
Inset in a): the JV sublattice with lattice parameters $a_J$ and
$b_J$ (lines mark the CuO$_2$ planes). Inset in b): the additional
deformation of the PV lattice which is required for trapping of
PVs by JVs. The upper sketch is the equilateral triangle of the
unperturbed PV lattice which is incommensurate with JV lattice
($a_J\ne pb$ with integer $p$). The lower sketch is the isosceles
triangle of the PV lattice matched with the JV sublattice
($a_J=p(b+\delta b)$).} \label{f3}
\end{center}
\end{figure}

\noindent 
Therefore, the ``trapped'' PV lattice seems to be realized as soon
as
 $a>\gamma s$.
In the opposite case, $\lambda_{ab}\gtrsim\gamma s$, the
transformation \cite{tran-sh-tr} from the ``shifted'' PV lattice
to the ``trapped'' PV lattice  occurs when the energy gain
$E_{tr}$ exceeds the energy loss $E_{shear}$. It happens in a
certain out-of-plane field between the field $\Phi_0/(\gamma
s)^2$, at which the ``crossing lattice pinning'' is activated, and
the field $B_z\sim \Phi_0/\lambda_{ab}^2$, where the shear elastic
energy rapidly decreases. Next, we discuss the difference between
the considered ``trapped'' state and the ``chain'' state proposed
for the crossing lattice \cite{koshelev}. The ``trapped'' state is
related to the rearrangement of PVs on the scale $a_J$ between the
nearest rows of JVs. On the other hand, the ``chain'' state is
associated with the creation of an extra PV row (an interstitial
in the PV lattice) on a JV, but the influence of the neighbouring
JVs is completely ignored. As a result, the ``trapped'' and
``chain'' states have the different
 in-plane field dependence of the out-of-plane transition fields.
 The out-of-plane
transition field \cite{struc} between the ``shifted'' and
``trapped'' PV lattices does not depend on $\Ha$ in contrast to
the $\Ha$-dependent out-of-plane field \cite{koshelev} of the
destruction of the ``chain'' state. Since the analysis
\cite{koshelev} is correct only in the case of $\gamma
s\gg\lambda_{ab}$ and $a\gg \lambda_{ab}$,
 the transformation of the PV lattice discussed here seems to be more likely
 in the case $\lambda_{ab}>\gamma s$.

\section{Phase diagram of vortex lattice in tilted
magnetic fields} \label{sec4}

In this section we discuss the vortex lattice structures formed at
 different field orientations. The tilted lattice consists
of mono-oriented vortices and transforms continuously from the
tilted PV stacks in fields near the $c$-axis (Fig. 4a) to the long
JV strings connected by PV kinks for the field orientations close
to the $ab$-plane (Fig. 4b). On the other hand, the tilted lattice
is topologically different from the crossing vortex structure
(Fig. 4c), and they replace each other via phase transition
\cite{ledvij}. For the analysis of the vortex
 phase diagram in tilted fields, the free energy of the crossing and tilted vortex
structures will be compared. We concentrate on the case $\gamma
s<\lambda_{ab}$, when, according to Bulaevskii {\it et al.}
\cite{ledvij} and Koshelev \cite{koshelev1}, the tilted lattice is
energetically preferable above the lock-in transition
\cite{lock-in}. We will consider a thin superconducting platelet
with the $c$-axis perpendicular to the plate. In this geometry the
lock-in transition occurs at very low fields \cite{ledvij}
$B_z\approx (1-n_z)\Phi_0/(4\pi\lambda_{ab}^2)\ln(\gamma
s/\xi_{ab})$ with demagnetization factor $n_z$ ($1-n_z\ll 1$).

For the field oriented close enough to the $c$-axis,
$\tan\theta=B_x/B_z\ll\gamma$, the free energy of the tilted
lattice $F_t$ can be evaluated as
$F_t=F_t^0+\frac{1}{2}C_{44}^{tilt}({\vec k}=0)B_x^2/B_z^2$ in
analogy to the analysis given in ref.~\cite{koshelev}. Here,
$F_t^0$ represents the free energy in the absence of the in-plane
magnetic field, while the tilt modulus is expressed as
$C_{44}^{tilt}({\vec k}=0)=B_z^2/4\pi + C_{44}^{eff}$ with
$C_{44}^{eff}$ defined in (\ref{u441}) for the case of
$B_z\gtrsim\Phi_0/(4\pi\lambda_{ab}^2)$. As a result, we have:

\begin{figure}[btp]
\begin{center}\leavevmode
\includegraphics[width=1.15\linewidth]{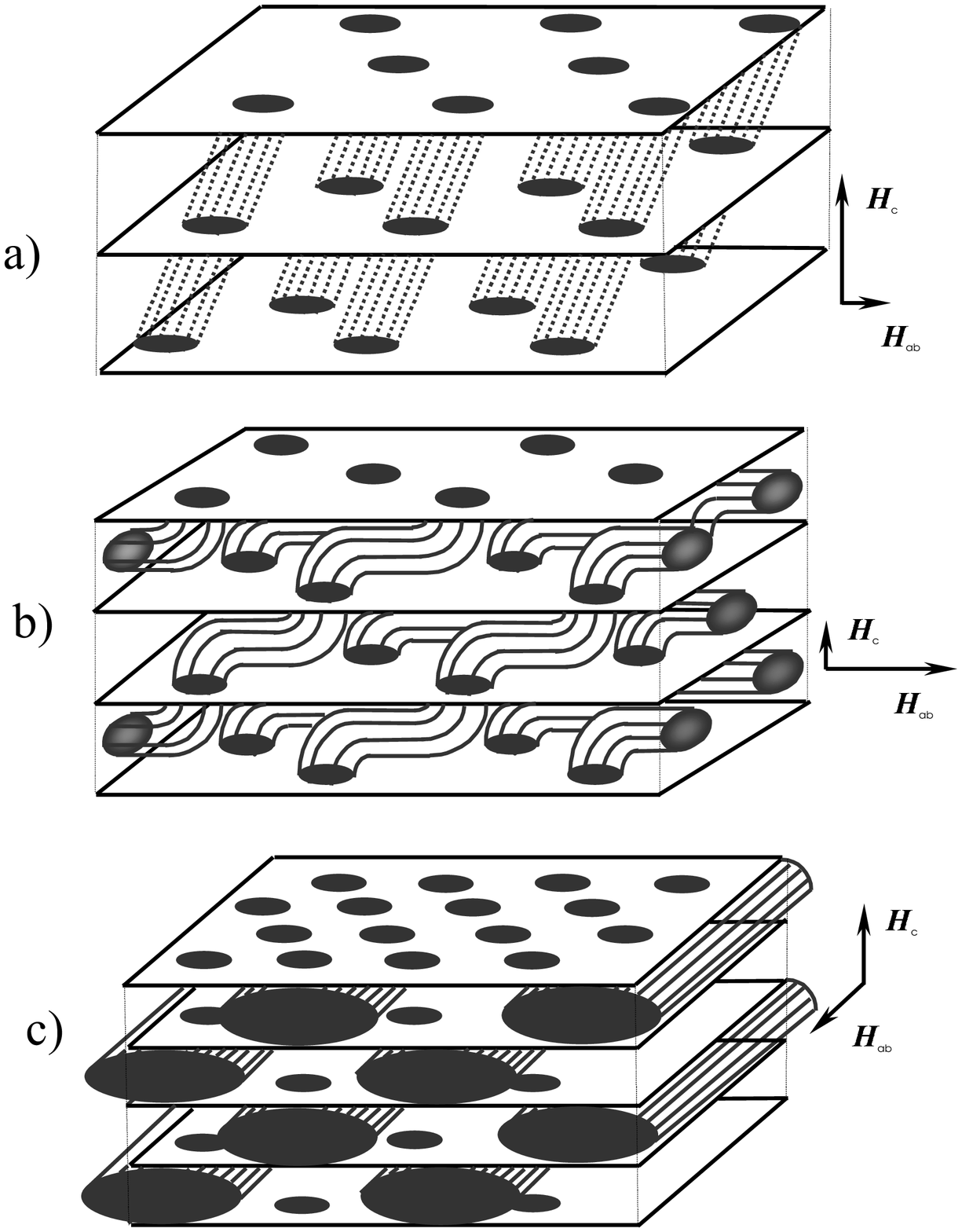}
\caption{The 3D sketches of the different vortex structures in the
tilted magnetic field with the components $H_c$ and $H_{ab}$ along
the $c$-axis and in the $ab$-plane, respectively: a) the tilted
vortex lattice near the $c$-axis (TI), when the current between
CuO$_2$ planes is much smaller than the critical value $j_c$, {\it
i. e.}, the Josephson strings linking PVs are not developed; b)
the tilted vortex lattice far away from the $c$-axis (TII), when
the JV strings are formed; c) the crossing vortex lattice.}
\label{f4}
\end{center}
\end{figure}

\begin{eqnarray}
F_{t}&\simeq &\frac{B_z^2}{8\pi}+\frac{\Phi_0 B_z}
{32\pi^2\lambda_{ab}^2}\ln{\frac{H_{c2\perp}}{B_z}}
+\frac{B_x^2}{8\pi} \nonumber \\ &+&
3.68\frac{\Phi_0^2}{2(4\pi\lambda_{ab})^4}\frac{B_x^2}{B_z^2}+
\frac{B_x^2\Phi_0} {64\pi^2\lambda_c^2
B_z}\ln\frac{H_{c2\perp}}{B_z},
\end{eqnarray}
where $H_{c2\perp}=\Phi_0/2\pi\xi_{ab}^2$. The first two terms
form the free energy for $B_x=0$. The third term is the in-plane
magnetic energy, the fourth one comes from the electromagnetic
interaction of the inclined PVs, and the last contribution is
connected with the Josephson coupling of PVs.

The free energy of the crossing lattice $F_c$ consists of two
contributions from the PV sublattice and the JV sublattice
while the interaction
of PVs and JVs is taken into account
 through the renormalization of the JV energy:
\begin{equation}
F_{c}\simeq \frac{B_z^2}{8\pi}+\frac{\Phi_0 B_z}
{32\pi^2\lambda_{ab}^2}\ln{\frac{H_{c2\perp}}{B_z}}+\frac{B_x^2}{8\pi}+
\frac{B_x}{\Phi_0}\E. \label{fc}
\end{equation}
The renormalized JV energy, $\E$, is defined by equation
(\ref{e0}) in which the lower limits of integrations are
restricted by the conditions $q_y,\ k_y\gtrsim 1/a_J$ and
$q_z,\ k_z\gtrsim 1/b_J$.

The tilted lattice is energetically preferable in the fields
oriented near the $c$-axis because $F_t\propto B_x^2$, while
$F_c\propto B_x$, {\it i.~e.}, $F_t<F_c$ for low $B_x$. The phase
boundary between the tilted lattice and the crossing structure can
be obtained from the condition $F_t=F_c$ which is rewritten in the
form:
\begin{equation}
B_x\simeq
\frac{\E}{\Phi_0}\frac{B_z^2}{1.84\Phi_0^2/(4\pi\lambda_{ab})^4+
\Phi_0B_z/(64\pi^2\lambda_c^2)\ln(H_{c2\perp}/B_z)}. \label{bound}
\end{equation}
The transition from the tilted lattice to the crossing structure
occurs at the field oriented quite close to the $c$-axis for high
anisotropic superconductors due to: 1) the high energy cost of the
inclination of PV stacks in the tilted lattice related to the
electromagnetic interaction of PVs, and 2) the decrease of the JV
energy in the crossing lattice structure.
For the dense PV lattice $B_z\gg\Phi_0/(\gamma s)^2$ and
$\lambda_{ab}>\gamma s$, equation (\ref{bound}) can be simplified:
\begin{equation}
B_x\simeq \sqrt{\frac{C_{66}}{\avu\lambda_{ab}^2}}
\frac{2\lambda_{ab}^2}{\lambda_J\lambda_{ab}^{eff}}\frac{B_z^2}{
\frac{\gamma\Phi_0}{4.3\pi^2\lambda_{ab}^2}+\frac{B_z}{2\gamma}\ln(H_{c2\perp}/B_z)}.
\end{equation}

Next, we will study the field orientations close to the
$ab$-plane, $B_x>\gamma B_z$. Here, the electromagnetic
interaction between PVs in the tilted lattice is not so important
and the free energy in the low $c$-axis fields $B_z<
\Phi_0/\lambda_{ab}^2$ is reduced \cite{ledvij} to:
\begin{equation}
F_t\simeq\frac{B_x^2}{8\pi}+\frac{\Phi_0B_x}{32\pi^2\lambda_{ab}\lambda_c}\ln\frac{\Phi_0}{\gamma
s^2 B_x} +\frac{H_JB_z}{4\pi}, \label{bul-tilt}
\end{equation}
where $H_J=\Phi_0/(4\pi\lambda_{ab}^2)\ln(\gamma s/\xi_{ab})$. The
first two terms are related to the energy of JV strings while the
last one is associated with the energy cost of the formation of PV
kinks. In more detail, the PV kink generates the in-plane current
which decreases with the distance $r$ from the kink center as
$1/r$  up to a critical radius $r_0$ of the region with 2D
behavior where the current along the $c$-axis is about the maximum
possible current $j_c$ \cite{ledvij}. At larger distances, the
in-plane current decays exponentially. Simple evaluation gives
$r_0=\gamma s$ for
 the PV kink \cite{ledvij}. We note, that the tilted vortex lattice
in the considered angular range of the magnetic field orientations
seems to exist as a kink-walls substructure, where kinks
(belonging to different vortices) are collected in separated walls
parallel to the $yz$-plane \cite{koshelev1}. For the kink-wall
substructure of the tilted lattice, the contribution to the free
energy (\ref{bul-tilt}), attributed to the PV kinks, is slightly
reduced in the high in-plane magnetic fields $B_x>\Phi_0/\gamma
s^2$ \cite{koshelev1}, which can be taken into account through
renormalization $H_J=\Phi_0/(8\pi\lambda_{ab}^2) \ln(\gamma
H_{c2}/B_x)$.

In the considered field interval, $B_x>\gamma B_z$,
$B_z\ll\Phi_0/\lambda_{ab}^2$, the renormalization of the JV
energy in the crossing lattice structure vanishes. However, the
interaction of PV and JV sublattices still manifests itself
through
 the ``crossing lattice pinning'':
\begin{eqnarray}
&&F_c=\frac{B_x^2}{8\pi}+
\frac{\Phi_0B_x}{32\pi^2\lambda_{ab}\lambda_c}\ln\frac{\Phi_0}{\gamma
s^2 B_x} +\frac{H_{c1\perp}B_z}{4\pi}
\nonumber \\
&&-\mu B_z\sqrt{
\frac{B_x\Phi_0}{16\pi^2\gamma\lambda_{ab}^2}}
\arctan\left(\frac{1-\sqrt{B_x/H_0}}{\sqrt{H_\lambda^x/H_0}+\sqrt{B_x/H_\lambda^x}}\right), \label{bul-cros}
\label{abfc}
\end{eqnarray}
with
$H_{c1\perp}=\Phi_0/(4\pi\lambda_{ab}^2)\ln(\lambda_{ab}/\xi_{ab})$
(the critical radius $r_0$ for the in-plane currents of a PV stack
is about $\lambda_{ab}$ \cite{ledvij}), $H_0=\Phi_0/\gamma s^2$
and $H_{\lambda}^x=\gamma\Phi_0/\lambda_{ab}^2$.
 The third term is the energy of the unperturbed PV lattice,
while the last term corresponds to the ``crossing lattice
pinning'' contribution which can significantly decrease the free
energy $F_c$ in the in-plane field interval $\gamma
\Phi_0/\lambda_{ab}^2\lesssim B_x\lesssim\Phi_0/\gamma s^2$. The
difference between the ``crossing lattice pinning'' contributions
to the free energy in the cases of low $B_x<\gamma B_z$ and high
$B_x>\gamma B_z$ in-plane fields (see equations (\ref{trap}),
(\ref{abfc})), emerges because the number of PV lines is
sufficient to occupy all JVs (Fig. 3b) at $B_x<\gamma B_z$ while
some JV strings do not carry PV rows (Fig. 3c) in the opposite
case. By analyzing equations (\ref{bul-tilt}) and (\ref{abfc}), we
can conclude that, at least for
$B_x\lesssim\gamma\Phi_0/\lambda_{ab}^2$ and
$B_x\gtrsim\Phi_0/\gamma s^2$, the tilted lattice exists near
the $ab$-plane since the condition $F_t<F_c$ is held
 due to the inequality $H_J<H_{c1\perp}$.
The tilted lattice is replaced by the crossing lattice with
increasing the out-of-plane magnetic field above $B_z=B_x/\gamma$.
However, it is difficult to determine the contour of the possible
phase line between the crossing and tilted vortex structures,
since it
 requires the more precise calculations of the free energies $F_c$ and
$F_t$ in the region $B_x<\gamma B_z$. In the intermediate in-plane
magnetic fields $\gamma\Phi_0/\lambda_{ab}^2 \lesssim
B_x\lesssim\Phi_0/\gamma s^2$, the ``crossing lattice pinning''
could make the crossing structure to be more energetically
preferable with respect to the tilted lattice. In that case, the
crossing lattice (CII, Fig. 3b) with $a<a_J$ transforms into the
crossing lattice (CIII, Fig. 3c) with the extremely dilute PV
sublattice $a>a_J$ at the angle $\theta=\arctan B_x/B_z\sim\arctan
\gamma$.

Therefore, we find a complicated picture of phase transitions
between the tilted vortex structure and the crossing vortex
structure
 in the case $\gamma s<\lambda_{ab}$.
The proposed phase diagram \cite{dem} is shown in figure 5. As it
was suggested earlier \cite{koshelev} and according to our
calculations by using equation (\ref{bound}), the tilted vortex
structure (TI) of inclined PV stacks (see Fig. 4a) can be replaced
by the crossing lattice quite close to the $c$-axis (see phase
diagram obtained for $\gamma=500$, inset in Fig. 5). 
\begin{figure}[btp]
\begin{center}\leavevmode
\includegraphics[width=1.15\linewidth]{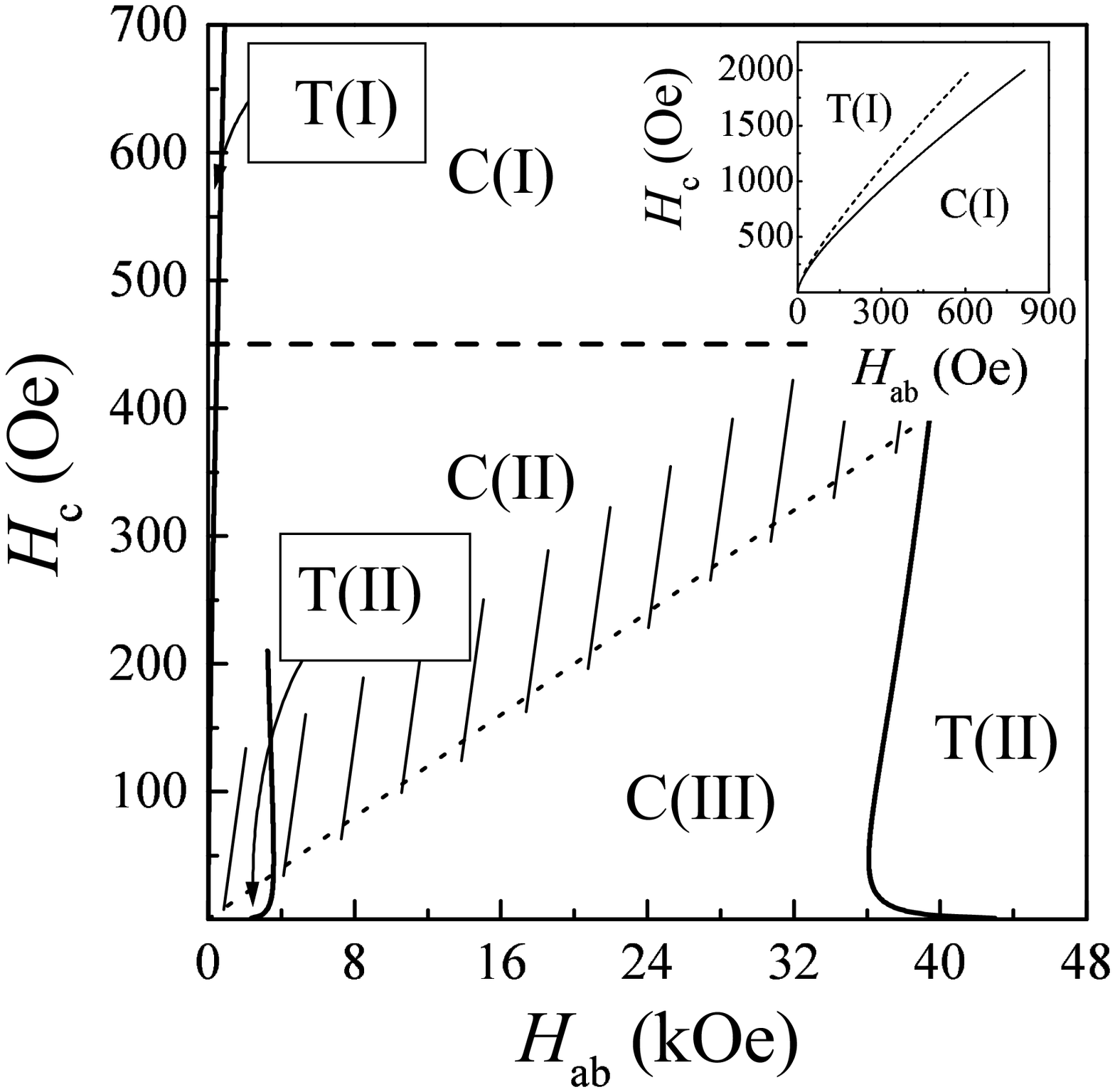}
\caption{The proposed phase diagram of the vortex solid phase in
the oblique magnetic fields calculated using equations
(\ref{trap}), (\ref{shear}), (\ref{bound}), (\ref{bul-tilt}) and
(\ref{abfc}) with parameters mentioned in the text, $T=45$ K,
$\gamma=100$ and $\nu=1$. The dotted line is the line $B_x=\gamma
B_z$, while the shaded area marks the region inside which the
transition from the crossing lattice C(II) to the tilted lattice T(II)
or the crossing lattice C(III)
happens. The arrows from enframed TI and TII are directed toward
the regions where these vortex structures are realized. Inset: the
part of the phase diagram close to the $c$-axis for strongly
anisotropic superconductors with $\gamma=500$ ($\gamma
s>\lambda_{ab}$) and $T=45$ K; the solid line marking the
transition from the tilted lattice (TI) to the crossing lattice is
obtained from eq.~(\ref{bound}), while the dashed line,
corresponding to the same transition, is calculated by using
eq.~(6) of Ref.{\protect{\cite{koshelev}}}. The parameters are
chosen to give some insight to the behavior of
Bi$_2$Sr$_2$CaCu$_2$O$_{8-\delta}$ in the oblique magnetic
fields.}
 \label{f5}
\end{center}
\end{figure}
\noindent Nevertheless,
in the high $c$-axis magnetic fields, the calculated in-plane
magnetic fields of this transition are higher than ones obtained
by using the model \cite{koshelev}. This difference comes from the
shear contribution
 to $\E$ which was omitted in ref.~\cite{koshelev}. The crossing lattice,
which exists in a wide angular range, can have different
substructures. At high enough out-of-plane fields
$B_z>\Phi_0/\gamma^2 s^2$, the ``shifted'' PV sublattice is
realized in the crossing lattice structure (CI, Fig. 3a). In this
substructure, the JV currents shift the PVs mostly along the
$x$-axis. The ``shifted'' phase can transform into the ``trapped''
PV lattice (CII, Fig. 3b), when the energy gain related to the
``crossing lattice pinning'' exceeds the energy needed for the
additional shear deformation (the dashed line in Fig. 5 separating
CI and CII has been obtained from the condition
$E_{tr}=E_{shear}$). Around the line $B_x=\gamma B_z$, the lattice
CII can be changed by the tilted lattice (TII) with JV strings
linked by PV kinks (Fig. 4b) or by the crossing lattice structure
(CIII, Fig. 3c) at which all PV stacks are placed on a few JVs.
The domain in the $H_c-H_{ab}$ phase diagram with the lattice CIII
is determined by the condition $F_c<F_t$ where $F_t$ and $F_c$ are
defined by the equations (\ref{bul-tilt}) and (\ref{abfc}),
respectively. With increasing temperature, the region of the
lattice CIII becomes narrower and disappears at a certain
temperature (see Fig. 6). The proposed phase diagram suggests the
possibility of the re-entrant tilted-crossing-tilted phase
transition as the magnetic field (at least with low
$B\lesssim\gamma\Phi_0/\lambda_{ab}^2$ or high $B\gtrsim
\Phi_0/\gamma s^2$ absolute value) is tilted away from the
$c$-axis to the $ab$-plane. Such possibility for low fields was
earlier mentioned in the works \cite{crossing-london,crossing-LD}
in which the interaction of crossed sublattices was not
considered. Moreover, we note that the instability of the tilted
lattice was found numerically by Thompson and Moore \cite{T-M}
(for $\gamma\lesssim 100$) only at the intermediate field
orientations $0^\circ<\theta_1<\theta<\theta_2<90^\circ$, which
could also support the discussed scenario. The parameters taken
for the $H_c-H_{ab}$ phase diagram at $T=45$ K (Fig. 5) and the
$H_{ab}-T$ phase diagram at the magnetic field orientation
$B_z=B_x/\gamma$ (Fig. 6) were chosen to give some insight into
the behavior of vortex array in BSCCO in the tilted magnetic
fields (see further discussion) as
$\lambda_{ab}=2000/\sqrt{1-T^2/T_c^2}\ {\rm \AA}$,
$\xi_{ab}=30/\sqrt{1-T/T_c}\ {\rm \AA}$, $s=15\ {\rm \AA}$,
$T_c=90$ K, $\gamma=100$ ($\gamma=500$ for inset in Fig.5 and
$\gamma=150$ for inset in Fig. 6).

\section{Conclusion}
\label{sec5}

This theoretical investigation was partially motivated by the
recent intensive experimental studies of the vortex lattice
melting transition \cite{ooi1,mir,we-prl,konc,jps} as well as
transitions in the vortex solid phase \cite{konc,ooi2} in $\rm
Bi_2Sr_2CaCu_2O_{8+\delta}$ single crystals. The observed linear
decay of the $c$-axis melting field component $\Hm$ with in-plane
fields \cite{ooi1,mir} was interpreted \cite{koshelev} as an
indication of the crossing vortex lattice. Thus, the tilted
lattice could be replaced by the crossing vortex structure quite
near the $c$-axis. 
According to our calculations, the angle where
such transition may occur is about 7$^\circ$ at $B_z\approx 100$
Oe and $\gamma\approx 500$ while that angle reaches 14.5$^\circ$
in the higher out-of-plane field $B_z=500$ Oe, which correlates
well with the disappearance of the hexagonal order along the
$c$-axis found by neutron measurements \cite{neutron}. 
With further tilting of the magnetic field, the linear dependence of
the $c$-axis melting field component $\Hm(\Ha)$ abruptly
transforms into a weak dependence \cite{mir,we-prl,konc}, which,
as was shown \cite{we-prl}, can not be explained in the frame of
the model \cite{koshelev}. Such behavior suggests a phase
transition in the vortex solid phase in tilted magnetic fields in $\rm
Bi_2Sr_2CaCu_2O_{8+\delta}$, which was detected in the recent ac
magnetization measurements \cite{konc,ooi2}. 
\begin{figure}[btp]
\begin{center}\leavevmode
\includegraphics[width=1.15\linewidth]{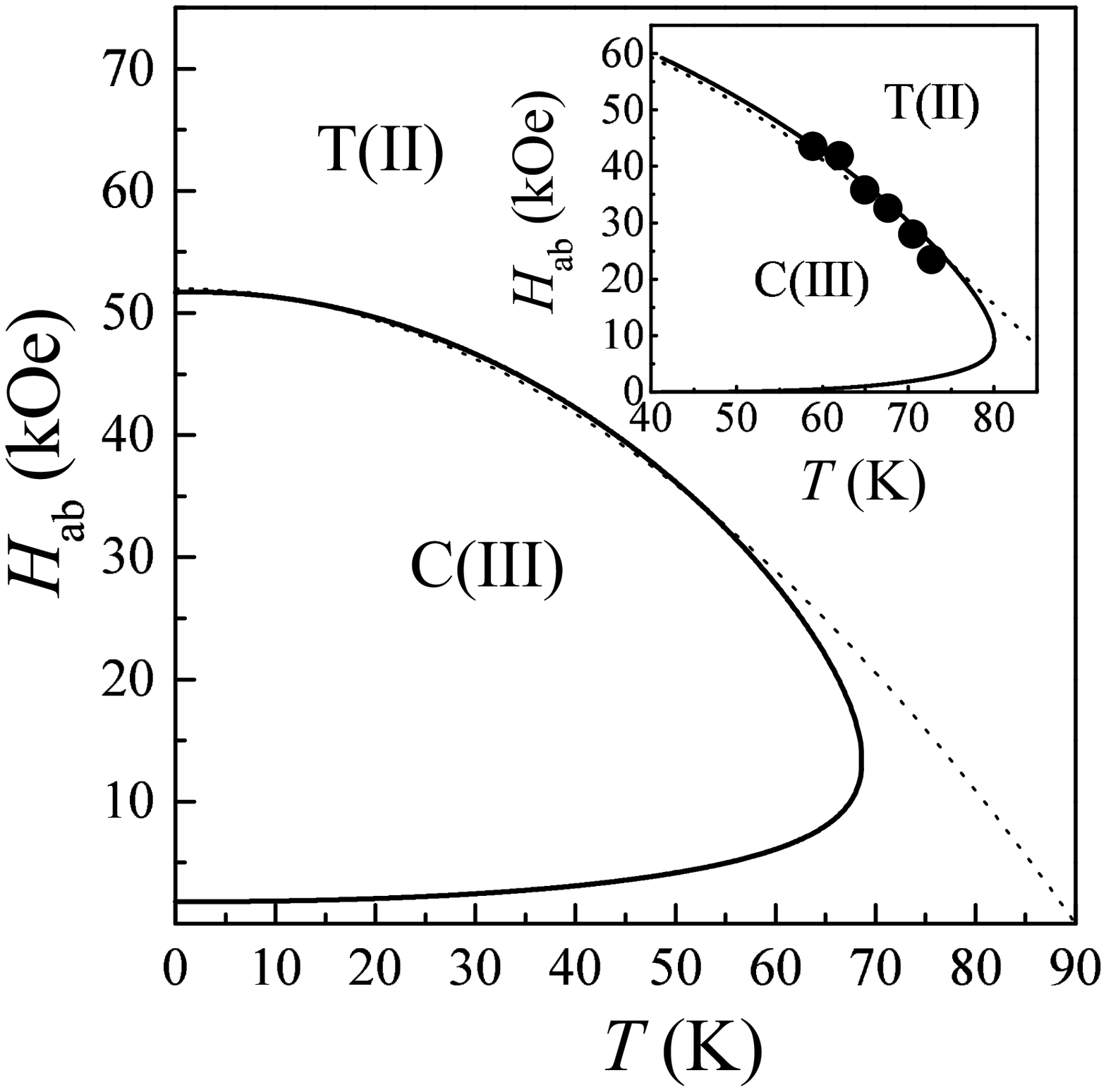}
\caption{The $H_{ab}-T$ phase diagram of the vortex solid phase at
the field oriented near the $ab$-plane ($B_x/B_z=\gamma=100$). The
region of the crossing lattice phase (CIII) becomes narrower and
finally disappears with increasing temperature. Inset: the same
phase diagram for $\gamma=150$. The dotted lines correspond to the
experimentally found temperature dependence $H_{ab}^m\propto
1-T^2/T_c^2$ of the in-plane characteristic fields of the vortex
lattice melting transition in Bi$_2$Sr$_2$CaCu$_2$O$_{8+\delta}$
in the tilted fields \protect{\cite{jps}}
 (for instance, symbols represented in the inset
exhibit the temperature dependence of the maximum in-plane field
of the vortex lattice melting transition).}
\label{f6}
\end{center}
\end{figure}
\noindent
As was mentioned by
Ooi {\it et al.} \cite{ooi2}, the behavior of the new anomaly of
the local magnetization in BSCCO attributed to the phase
transition in the vortex solid slightly reminds the peak effect
related to the vortex pinning, which, in turn, could be induced by
the vortex trapping by planar defects \cite{rakh}. Such analogy,
as well as a very weak dependence of the $c$-axis magnetic field component
$H_c^*$ of the novel phase transition on in-plane magnetic fields
\cite{ooi2} in a wide angular range, could suggest the tansition
from the ``shifted'' PV sublattice (CI) to the ``trapped'' PV
sublattice (CII) in the crossing lattice structure. Our estimation
of the $c$-axis field $H_c^{trap}$ of the transition from CI to
CII ($H_c^{trap}\approx 450$ Oe for $\gamma=100$ at $T=45$ K (see
Fig. 5)) \cite{razbros} is in a reasonable agreement with
experimental findings \cite{ooi2} $H_c^*(T=45\ {\rm K})\approx
430$ Oe. Near the $ab$-plane, the properties of the observed
anomaly changes abruptly and the field $H_c^*$ sharply goes to
zero \cite{ooi2}, which may indicate a transformation \cite{struc}
in the JV lattice or a trace of the phase transition from TII to
CIII. At higher in-plane fields, the found step-wise behavior
\cite{we-prl} of the vortex lattice melting transition may be
related to the existence of one more phase transition in the solid
phase. Interestingly enough, all characteristic in-plane fields of
the vortex lattice melting transition depend on temperature
\cite{jps} proportionally to $1-T^2/T_c^2$, which is similar to
the calculated temperature dependence of the phase transition from
CIII to TII (see Fig. 6).

In summary, we discussed the crossing lattice structure in a strongly
anisotropic layered superconductor in the framework of the
extended anisotropic London theory. The renormalization of the JV
energy in the crossing lattice structure was calculated in the
cases of the dense PV lattice as well as the dilute PV lattice. It
was shown, that the ``crossing lattice pinning'' can induce the
rearrangement of the PV sublattice in the crossing lattice
structure as soon as the out-of-plane magnetic field becomes lower
than a certain critical value. The free energy analysis indicates
a possibility of the re-entrant tilted-crossing-tilted lattice
phase transition with inclination of the magnetic field away from
the $c$-axis to the $ab$-plane in the case of $\lambda_{ab}>\gamma
s$.

{\renewcommand{\theequation}{{\rm A}\arabic{equation}}
\setcounter{equation}{0}

\newcommand{\FCR}{(\ref{f-cr-k})}
\section*{Appendix A. Approximate summation in equation \FCR}
Our aim is to sum the sequence:
\begin{eqnarray}
&\Phi_0&\sum_{n=-\infty}^{\infty}ik_zu({\vec k})
\frac{f(k_z,k_y-2\pi n/b)-(B_z/2\Phi_0)ik_zu(-{\vec k})}{1+\lambda_{ab}^2k_z^2+
\lambda_c^2(k_y-2\pi n/b)^2}= \nonumber\\
&ik_z&\Psi_1(k_y,k_z)u({\vec k})+
(B_z/2\Phi_0)k_z^2\Psi(k_y,k_z)u({\vec k})u(-{\vec k}).
\end{eqnarray}
The equation (\ref{psi0}) for $\Psi(k_y,k_z)$ can be directly obtained by
using the well-known mathematical equality
$$
\sum_{n}\frac{1}{r^2+(t+2\pi n/\alpha)^2}=\frac{\alpha}{2r}
\frac{\sh(r\alpha)}{\ch(r\alpha)-\cos(t\alpha)}
$$
with real numbers $r,\ t,$ and $\alpha$.

Next the sum $\Psi_1(k_y,k_z)=\Phi_0\sum_{Q_y=2\pi n/b}f(k_z,k_y-Q_y)/(1+\lambda_{ab}^2
k_z^2+\lambda_c^2(k_y-Q_y)^2)$ needs to be estimated. By using inequailty
$k_z\lesssim 1/s$, we obtain $f(k_z,k_y-Q_y)\approx f_J(k_y-Q_y)$ where
$f_J\approx 1$ for $|k_y-Q_y|\lesssim 1/\lambda_J$ and zero otherwise.
In the case of the dense PV lattice ($b\ll\lambda_J$) we retain only the term
with $n=0$ in the sum and get the expression $\Psi_1(k_y>1/\lambda_J)=0$
and $\Psi_1(k_y\lesssim 1/\lambda_J)=
\Phi_0/(1+\lambda_{ab}^2k_z^2+\lambda_c^2k_y^2)$. Taking into account the
inequality $k_y<1/\lambda_J\ll 1/b$ and
$\sqrt{1+\lambda_{ab}^2k_z^2}b/\lambda_c<b/(\gamma s)\ll 1$,
one can rewrite $\Psi_1(k_y<1/\lambda_J,b\ll\lambda_J)\simeq
\Psi(k_y,k_z)$. Thus, we come to the equation (\ref{f-cr-last}).

For the case of the dilute PV lattice ($\lambda_J\ll b$),
many terms give contributions
to the sum $\Psi_1\simeq\Phi_0\sum_{n=-N}^{N}1/(1+\lambda_{ab}^2k_z^2+
\lambda_c^2(k_y-2\pi n/b)^2)$, since inequality $|k_y-2\pi n/b|<1/\lambda_J$
is held until $n$ exceeds $N\gg 1$. Thus, the function $\Psi_1$ is estimated
as:
\begin{equation}
\Psi_1(k_y,k_z)\approx\Psi(k_y,k_z)-
\frac{\Phi_0b}{\pi}\int_{1/\lambda_J}^{\infty}\frac{dx}{1+\lambda_{ab}^2k_z^2+
\lambda_c^2x^2}
\end{equation}
Finally, we obtain $\Psi_1=(1-\beta(k_y,k_z))\Psi$ with
\begin{eqnarray}
\beta(k_y,k_z)\approx
&&\frac{\ch(\sqrt{1+\lambda_{ab}^2k_z^2}b/\lambda_c)-\cos(k_yb)}
{\sh(\sqrt{1+\lambda_{ab}^2k_z^2}b/\lambda_c)} \nonumber \\ &&\left(1-\frac{2}{\pi}
\arctan\left(\frac{\lambda_c}{\lambda_J\sqrt{1+\lambda_{ab}^2k_z^2}}\right)
\right)
\end{eqnarray}
At $|k_z|\ll 1/s$, it is easy to show that $\beta\lesssim \lambda_J/b\ll 1$
and the approximation $\Psi_1=\Psi$ used in (\ref{f-cr-last}) is excellent.
Only for $k_z\sim 1/s$, the function $\Psi_1$ can differ from the function
$\Psi$ by a factor about unity in the case of the dilute PV
lattice (the factor is $1-\beta\approx 0.5$ in the framework
of our rough consideration). However, the correct estimation of the
value of the factor depends on the type of the smoothing function and requires
more precise analisys than one in the framework of the London approach.
Therefore, we can always assume $\Psi_1=\Psi$ in our
semiquantitative consideration.}

{\renewcommand{\theequation}{{\rm B}\arabic{equation}}

\newcommand{\REFI}{\ref{ej-a<lambdaJ}}
\newcommand{\REFII}{\ref{lambdaJ}}
\newcommand{\REFIII}{\ref{e+pin}}

\setcounter{equation}{0}
\section*{Appendix B. Evaluation of integrals in equations (\REFI,\REFII,\REFIII)}

In this appendix, the integrals in equations
(\ref{ej-a<lambdaJ},\ref{lambdaJ},\ref{e+pin}) are evaluated. We
start with the dense PV lattice, $a\ll\lambda_J$. In the
region $q_z<1/b$, the tilt energy is small (\ref{u441}). Therefore, the
denominators of the integrands in (\ref{ej-a<lambdaJ},\ref{lambdaJ}) are
substantially larger than ones in the case $q_z>1/b$ and we can roughly
neglect the contribution related to the region $q_z<1/b$. Using equation
(\ref{uav}) for the tilt energy in the domain $q_z>1/b$, the integrals
(\ref{ej-a<lambdaJ},\ref{lambdaJ}) can be rewritten as follows:

\newpage
{
\twocolumn[\hsize\textwidth\columnwidth\hsize\csname@twocolumnfalse\endcsname
\begin{eqnarray}
I=2\int_{-1/\lambda_J}^{1/\lambda_J}dq_y\int_{1/b}^{1/s}dq_z \cdot
\frac{f(q_y^2)}{
1+\lambda_c^2q_y^2+\lambda_{ab}^2q_z^2+B_z^2q_z^2/(4\pi(\avu+\avc
q_z^2+C_{66}q_y^2))}, \nonumber \\ \label{i1}
\end{eqnarray}
where $f(q_y^2)=\Phi_0^2/32\pi^3$ for (\ref{ej-a<lambdaJ}) and
$f=\lambda_J q_y^2$ for (\ref{lambdaJ}). After multiplying
numerator and denominator by factor of $\avu+\avc
q_z^2+C_{66}q_y^2$, the expression (\ref{i1}) is reduced to
\begin{eqnarray}
I=2\int_{-1/\lambda_J}^{1/\lambda_J}dq_y\int_{1/b}^{1/s}dq_z \cdot
\frac{(\avu+\avc q_z^2+C_{66}q_y^2)f(q_y^2)}
{(1+\lambda_c^2q_y^2+\lambda_{ab}^2q_z^2)(\avu+C_{66}q_y^2)+
\Bigl((1+\lambda_c^2q_y^2+\lambda_{ab}^2q_z^2)
\avc+\frac{B_z^2}{4\pi}\Bigr)q_z^2} \label{i00}
\end{eqnarray}
The term $(1+\lambda_c^2q_y^2+\lambda_{ab}^2q_z^2)\avc$ can be
neglected with respect to $B_z^2/4\pi$ in the denominator. Indeed,
the maximum value of $\lambda_c^2q_y^2\avc$ (see eq.
(\ref{uav})) is about $\lambda_c^2\avc 1/\lambda_J^2\sim
B_z\Phi_0/\lambda_J^2\ll B_z^2$, while the maximum value of
$\lambda_{ab}^2q_z^2\avc$ is $\lambda_{c}^2\avc
1/\lambda_J^2\times(\lambda_J/\gamma s)^2$ which is even smaller,
since $\lambda_J<\gamma s$ (see eq.~(\ref{lambdaJ-K})). Next, the
integration in (\ref{i00}) over $q_z$ can be taken easily:
\begin{eqnarray}
I=4\int_{1/\lambda_c}^{1/\lambda_J} d q_y
f(q_y^2)\Biggl(\frac{\avc}{B_z^2/4\pi+\lambda_{ab}^2
(\avu+C_{66}q_y^2)}\frac{1}{s}+
\frac{1}{\lambda_{c}q_y}\frac{\sqrt{\avu+C_{66}q_y^2}\theta^*(q_y)}
{\sqrt{B_z^2/4\pi+\lambda_{ab}^2(\avu+C_{66}q_y^2)}}\Biggr)
\label{i01}
\end{eqnarray}
where the function $\theta^*$, defined as
$$\theta^*(q_y)=\arctan\left(\frac{q_y^{-1}}{\gamma s}
\sqrt{1+B_z^2/(4\pi\lambda_{ab}^2(\avu+C_{66}q_y^2))}\right)
- \arctan\left(\frac{q_y^{-1}}{\gamma b}
\sqrt{1+B_z^2/(4\pi\lambda_{ab}^2(\avu+C_{66}q_y^2))}\right),$$
 determines the lower cutting value $1/\lambda_{cut}$
of $q_y$. Namely, by using (\ref{lambdaJ-K}) we can assume
that the argument of the first arc tangent in
the last expression is larger than 1, for most of values of $q_y$.
Then, the value of
$\theta^*(q_y)$ is about $\pi/2$ if the argument of the second
 arc tangent is smaller than 1; otherwise $\theta^*$ is close to
zero. Thus, we obtain the expression for $\lambda_{cut}$:
$$
\lambda_{cut}\approx
\frac{b\gamma}{\sqrt{1+\frac{B_z^2}{4\pi\lambda_{ab}^2
(\avu+C_{66}\lambda_{cut}^{-2})}}}
\approx
\frac{\sqrt{2}b\gamma}{\left(1-\frac{b^2\gamma^2}{\delta^2}+\sqrt{\left(\frac{b^2\gamma^2}{\delta^2}-1\right)^2
+4\frac{(\lambda_{ab}^{eff})^2b^2\gamma^2}{\lambda_{ab}^2\delta^2}}\right)^{1/2}}
$$
$$
\sim \min(\lambda_{c},\min(\gamma b,
\max(b\lambda_c/\lambda_{ab}^{eff},\sqrt{b\lambda_c\delta/\lambda_{ab}^{eff}}))),
$$
where we denote $\delta=\sqrt{C_{66}/\avu}\sim \lambda_{ab}$ and
take into account that $\lambda_{cut}$ should be smaller than
 $\lambda_{c}$.
 As a result, the integral (\ref{i01}) can be evaluated
as:
\begin{eqnarray}
I=4\int_{1/\lambda_c}^{1/\lambda_J}  d q_y f(q_y^2)\frac{\avc}{B_z^2/4\pi+\lambda_{ab}^2
(\avu+C_{66}q_y^2)}\frac{1}{s}+
2\pi\int_{1/\lambda_{cut}}^{1/\lambda_J}  d q_y f(q_y^2)
\frac{1}{\lambda_{c}q_y}\frac{\sqrt{\avu+C_{66}
q_y^2}}
{\sqrt{B_z^2/4\pi+\lambda_{ab}^2(\avu+C_{66}q_y^2)}}. \label{I-general}
\end{eqnarray}
In case of the dense PV lattice, the first integral is
small and can be neglected. 
In addition,
 we also can omit the term $C_{66}q_y^2$ in the
expression $B_z^2/4\pi +\lambda_{ab}^2C_{66}q_y^2$. Finally, we
have
\begin{equation}
I(a\ll\lambda_{J})=2\pi\frac{1}{\lambda_c\lambda_{ab}^{eff}}\int_{1/\lambda_{cut}}^{1/\lambda_{J}}
\frac{f(q_y^2)dq_y}{q_y}\sqrt{1+\frac{C_{66}}{\avu}q_y^2}.
\label{I-dense}
\end{equation}
\vskip.2pc]} 

\
\newpage
\
\newpage
After taking $f(q_y^2)=\lambda_J q_y^2$ in (\ref{I-dense}) and by
ignoring $C_{66}q_y^2/\avu$ in the case $\lambda_J>\lambda_{ab}$
or by neglecting unity in the opposite case, we obtain the
expression
$\lambda_{J}\approx\max{\left(\frac{\lambda_cs}{\lambda_{ab}^{eff}},\sqrt{\frac{\lambda_cs}{
\lambda_{ab}^{eff}}\delta}\right)}$. 
It coincides with
(\ref{lambdaJ-K}) since $\delta\approx \lambda_{ab}$. The energy
of JV (\ref{ej-k}) is easily derived if one puts
$f(q_y^2)=\Phi_0^2/(32\pi^3)$ in (\ref{I-dense}). \break 
Next, we roughly estimate the first integral in equation
(\ref{e+pin}). The inequality
$(1+\lambda_c^2q_y^2+\lambda_{ab}^2q_z^2)\avc\ll B_z^2$ is still
correct in the domain $q_z\ll \gamma/b$, $q_y\ll 1/b$ for
$B_z\gg\Phi_0/\lambda_c^2$. Thus, we can get the estimation
(\ref{I-general}) also for the dilute case ($a>\gamma s$).
However, in contrary to the dense PV lattice, the contribution
related to the first term in (\ref{I-general}) remains important:
\begin{eqnarray}
I(a\gg\lambda_{J})&=&4\frac{\avc\gamma}{\avu
b(\lambda_{ab}^{eff})^2}\int_{1/\lambda_c}^{\mu_1/b}dq_y f(q_y^2)
\nonumber \\ &+&\frac{2\pi}{\lambda_{c}\lambda_{ab}^{eff}}\int_{1/\lambda_{cut}}^{\mu_2/b}dq_y
\frac{f(q_y^2)}{q_y}\sqrt{1+\frac{C_{66}}{\avu}q_y^2}.
\end{eqnarray}
Here, we have introduced numerical parameters $\mu_1$ and $\mu_2$,
since the upper limits of integration are not well defined. The
corresponding contribution to the energy is obtained from the last
equation for $f=\Phi_0^2/32\pi^3$ as presented in the text.
}


\begin{references}
\bibitem{tran-sh-tr} One of the possible physical pictures of
the replacement of the ``shifted'' PV lattice by the ``trapped''
PV lattice is the sequence of the phase transitions. At each of
those transitions, the distance $L$ between the nearest JV rows,
occupied by PV rows, increases by $a_J$ and takes values $L=a_J$
(lattice in Fig. 3b), $L=2a_J$ (every second JV row is occupied),
etc.
\bibitem{struc} The structural transitions between the Josephson
vortex lattices with different periods \cite{bul},
which are related to the layerness
of the medium, cause
the nonmonotonic dependence of $a_J$ on the in-plane
field \cite{we-prl}. Such behavior of the JV sublattice can induce the
complicated alternation of the ``shifted'' and ``trapped'' states.
\end{references}
\end{document}